\documentclass[a4paper,11pt]{article}

\usepackage{times}
\usepackage{amssymb}
\usepackage{epsfig}

\newcommand{\beq}{\begin{equation}}
\newcommand{\eeq}{\end{equation}}
\newcommand{\beqa}{\begin{eqnarray}}
\newcommand{\eeqa}{\end{eqnarray}}
\newcommand{\beqas}{\begin{eqnarray*}}
\newcommand{\eeqas}{\end{eqnarray*}}

\newcommand{\ceq}{\!\!\! & = & \!\!\!}
\newcommand{\cleq}{\!\!\! & \le & \!\!\!}

\newcommand{\done}{\hfill $\Box$}

\newcommand{\real}{{\ensuremath{\mathbb{R}}}}

\newcommand{\dx}{{\dot x}}
\newcommand{\pd}{{\partial}}

\newcommand{\llangle}{{\bigg\langle}}
\newcommand{\rrangle}{{\bigg\rangle}}

\newcommand{\X}{\mathcal{X}}
\newcommand{\Fp}{\mathfrak{p}}
\newcommand{\Fn}{\mathfrak{n}}
\newcommand{\Ft}{\mathfrak{t}}
\newcommand{\FC}{\mathfrak{C}}

\newtheorem{prop}{Proposition}
\newtheorem{lem}{Lemma}
\newtheorem{rem}{Remark}

\newcommand{\eg}{e.\,g.}
\newcommand{\ie}{i.\,e.}
\newcommand{\wrt}{w.\,r.\,t.}

\newcommand{\LCD}{{\sc lcd}}
\newcommand{\GOI}{{\sc goi}}
\newcommand{\PDF}{{\sc pdf}}

\newcommand{\VP}{{\sc vp}}
\newcommand{\GP}{{\sc gp}}

\newcommand{\mytitle}[4]{%
\begin{center}
\LARGE #1
\par\bigskip
\normalsize\rm #2
\par\bigskip
\small #3
\par\medskip
\small #4
\end{center}
}
\newcommand{\fns}{\footnotesize}
\newcommand{\insfig}[2]{\epsfig{file=#1.eps,scale=#2}}
\newcommand{\insfigc}[2]{\centerline{\insfig{#1}{#2}}}
\newcommand{\babs}{\begin{quote}\small{\bf Abstract} --- }
\newcommand{\eabs}{\end{quote}}

\newcommand{\book}[5]{{#1} (#5). {\it #2}. {#4}: {#3}.}
\newcommand{\jour}[6]{{#1} (#6). {#2} {\it #3}, {\it #4}, {#5}.}
\newcommand{\proc}[8]{%
{#1} (#8). {#2} In #3 (Eds.), {\it #4} (pp. #7). {#6}: {#5}.}

\begin{document}
\mytitle
{Modeling geometric--optical illusions:\\A variational approach}
{Werner Ehm and Ji\v{r}\'i Wackermann}
{Institute for Frontier Areas of Psychology and Mental Health,\\Freiburg, Germany}
{\{ehm,\,jw\}@igpp.de}

\babs
Visual distortions of perceived lengths, angles, or forms, are generally known as ``geometric--optical illusions'' (\GOI). In the present paper we focus on a class of {\GOI}s where the distortion of a straight line segment (the ``target'' stimulus) is induced by an array of non-intersecting curvilinear elements (``context'' stimulus). Assuming local target--context interactions in a vector field representation of the context, we propose to model the perceptual distortion of the target as the solution to a minimization problem in the calculus of variations.  We discuss properties of the solutions and reproduction of the respective form of the perceptual distortion for several types of contexts. Moreover, we draw a connection between the interactionist model of {\GOI}s and Riemannian geometry: the context stimulus is understood as perturbing the geometry of the visual field from which the illusory distortion naturally arises. The approach is illustrated by data from a psychophysical experiment with nine subjects and six different contexts.  

\eabs

\section{Introduction}\label{intro}

``Geometric-optical illusions'' (\GOI) is a covering term for a broad class of phenomena, where visual perception of lengths, angles, areas or forms in a figure (\eg\ a simple line drawing) is altered by other components of the figure. These phenomena demonstrate, generally, the dependence of a percept on its context, and allow to study the structural principles underlying the organization of visual percepts, or ``laws of seeing'' \cite{Met75}. Since their discovery \cite{Opp55,Opp61}, {\GOI}s have been the subject of intensive experimental research (for comprehensive reviews see \cite{CoGi78} and \cite{Rob98}), but they are still far from being well understood. The variety of proposed explanations ranges from physiological theories, based on mutual interactions between elements of the neural substrate (\eg, retina or primary cortical areas) \cite{Bek67,CaBl73,Wal73}, to purely mentalist theories, interpreting the {\GOI}s as results of ``unconscious inferences'' \cite{Hel67} or inappropriately applied cognitive strategies \cite{Gre63}. However, no unitary theory of the {\GOI}s has been established until present days, and it is even doubtful whether such a unified explanatory theory is conceivable \cite{CoGi78}.

In the present paper we study a well-defined class of {\GOI}s that are reducible to a common generating principle. The emphasis is not on the human vision system or on psychological factors, nor will physiological or psychological ``mechanisms'' be proposed; we aim at a representative-descriptive rather than explanatory-causal theory. 
Specifically, we focus on a class of {\GOI}s in which perception of a {\em target} element---usually a segment of a straight line---appears distorted when presented with an array of (curvi)linear elements, in the following called the {\em context}. An example for such target--context interactions first reported by Hering \cite{Her61} is the illusory curvature of straight lines over which an array of concurrent lines is superposed (Fig.~1a). Since then a great number of {\GOI}s have been constructed, discovered, or re-discovered on the same principle \cite{Wun98,Ehr25,Orb39} (Figs.~1b,d,e).

\begin{figure}[t]
\insfigc{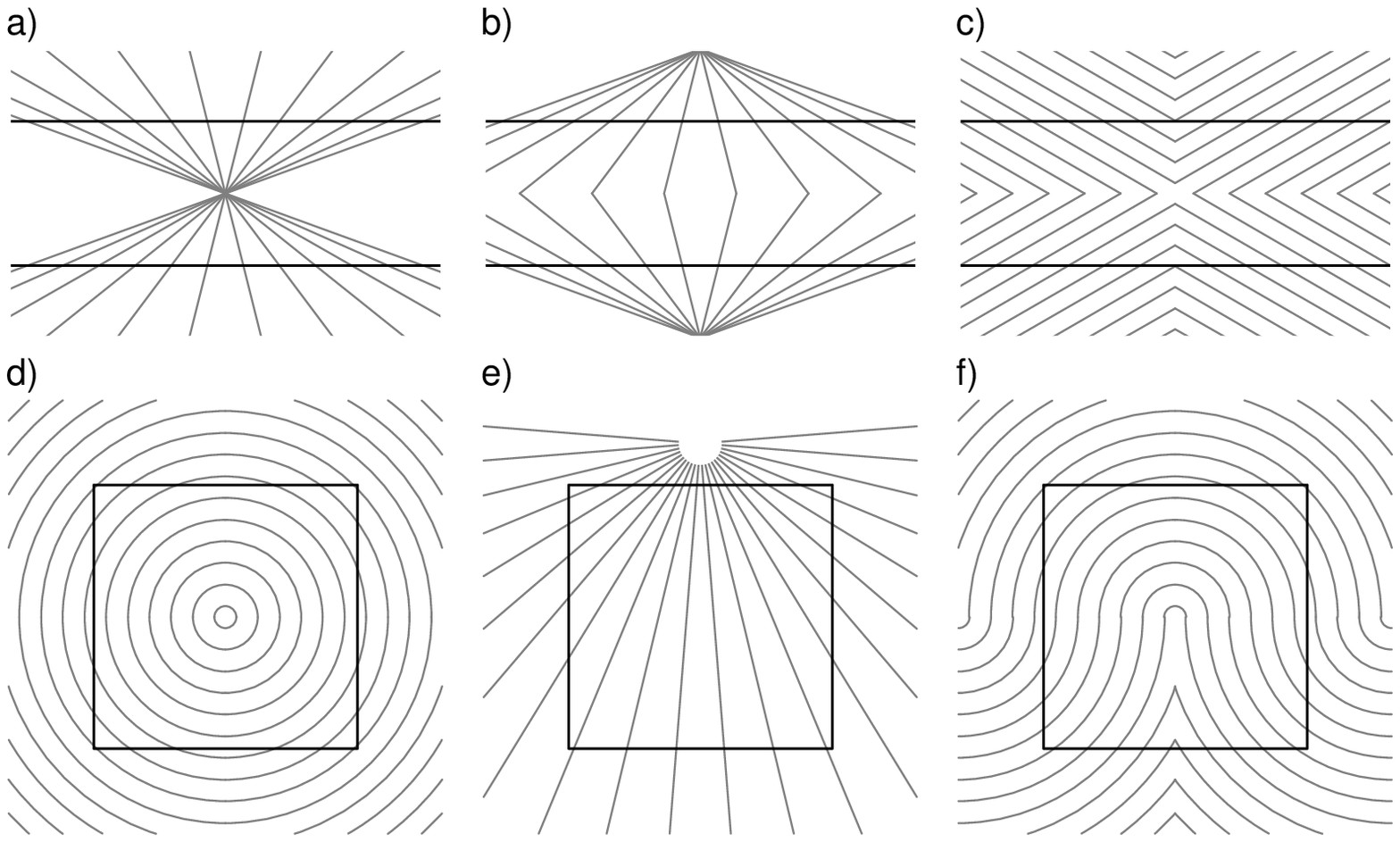}{0.875}

\medskip
\fns
Figure~1: Examples of geometric--optical illusions. Upper row: a)~Classic form of Hering's \cite{Her61} illusion and b)~its modification due to Thi\'ery and Wundt \cite{Wun98}; c)~Illusory bending of straight lines in a flat, non-perspectival context. Lower row: d,e)~Distortions of square shape in two different contexts \cite{Ehr25,Orb39}; f)~Trapezoid deformation of square shape similar to e) in a~different context, obtained by permutation of quadrants of pattern~d) \cite{Wac10}.
\end{figure}

\normalsize
These phenomena---hereafter called illusions of ``Hering type''---are of particular interest for several reasons. First, they depend on {\em local} interactions between the target and the context elements, as is evidenced by variant figures in which parts of the context pattern are deleted \cite{Wac10}. Next, they demonstrably do {\em not} depend on a ``scenic'' impression induced by the context patterns (Fig.~1e,f). Finally, they all exhibit {\em angular expansion} at the target--context intersections: the illusory distortion of the target always acts to enlarge the acute angles at the intersection points (Fig.~1, passim).  This effect, also dubbed ``regression to right angles,'' seems to be constitutive for the class of {\GOI}s of our interest \cite{BeBe48,Hor71} as well as in other types of {\GOI}s \cite{HoOl72,HoRo75}.

These observations set up the framework for our modeling approach \cite{Ehm10}. Starting with a minimal set of assumptions plus the fact that the straight line is the shortest path connecting two points, we propose a variational principle for the perception of a linear target, draw a connection to Riemannian geometry, and show that approximate solutions of the respective variational problem reproduce the perceptual distortions of the target (Sections~\ref{VP} to \ref{exs}). Further, we report on a related psycho--physical pilot experiment using six different context patterns (Section~\ref{exp}). Finally, we discuss achievements and limitations of this work (Section~\ref{disc}). 

The main text covers the basic approach along with the applied methods and the results. All mathematical details, derivations, and proofs are given in the Appendix.

\section{Variational problem for Hering type illusions}\label{VP}

Our focus in this paper is on the case where the target is a straight line, and the context consists of a family of (generally curved) lines that intersect the target but not each other. We will conceive of the context lines as the stream lines of a planar flow given by a continuously differentiable vector field $v$ defined on some region $\Xi \subseteq \real^2$ containing the target in its interior.\footnote{%
For clarity it has to be emphasized that the vector field $v$ just serves us to represent the context; it is neither related to any kind of field theory or perceptive field, nor to the receptive field of the retina. Likewise, there is no supposition as to where the percept is located ``materially.'' Finally, the term ``flow'' is used only metaphorically; it shall {\em not} convey any idea of motion. }
To any point $\xi \in \Xi$ is attached a vector, $v(\xi)$, indicating the ``velocity'' of the flow at the point $\xi$. In view of the purely geometric character of the context it is natural to assume that $|v(\xi)| = 1$ for all $\xi \in \Xi$. Here $|a| = \sqrt{\langle a, a\rangle}$ and $\langle a, b\rangle = a_1 b_1 + a_2 b_2$ denote the Euclidean norm (length) and inner product, respectively, of vectors $a,b \in \real^2$. In geometrical terms, the normalized inner product $\langle a, b\rangle/ |a||b|$ gives the cosine of the angle between $a$ and $b$, which we denote as $\angle (a,b)$. 

In graphical presentations of {\GOI}s only a finite sample of context curves is displayed. The complete set of context lines, which form a continuum in the plane, may then be conceived of as continuously interpolating the sample. The target is here assumed to be the straight line, $\tau$, connecting two given endpoints $\tau_0,\, \tau_1 \in \real^2$.\footnote{Subscripts may index objects (such as $\tau_0,\, \tau_1$) as well as the components of a vector (such that $y = (y_1, y_2)$ if $y \in \real^2$ is a row vector). The appropriate interpretation will always be evident from the context.} In illusions of Hering type $\tau$ is not perceived as a straight line: it appears slightly curved. The basic idea of our approach is to model the deviating percept as a perturbation of $\tau$ that is characterized by a minimum principle. Setting up the principle involves three components:

\vspace{-1ex}
\begin{enumerate}
\setlength{\itemsep}{0pt}
\item[(a)]
the local interactions hypothesis: the context $v$ ``acts'' only along candidate paths, in the vicinity of the target;
\item[(b)]
the angular expansion hypothesis (``regression to right angles''), based on the phenomenology of {\GOI}s (cf.\ Introduction);
\item[(c)]
the fact that the straight line is the shortest path between two points.
\end{enumerate}

\vspace{-1ex}
\noindent
Observing (b) and (c) we then posit the principle that, {\em given the context vector field $v$, the straight line target $\tau$ is distorted so that (i)~the stream lines of $v$ (the context lines) are intersected ``as orthogonally as possible'', and (ii)~the distorted line is as short as possible.}

This can be formulated mathematically as an optimization problem under side conditions. Since there is no {\em a~priori} criterion suggesting length or orthogonality as the primary or the side condition, we propose to optimize a weighted mixture of the two terms. Specifically, we consider the following 

\medskip\noindent
{\bf Variational problem [\VP]: } {\em Given the vector field $v$, $t_0,\, t_1 \in \real$ such that $t_0 < t_1$, endpoints $\tau_0$, $\tau_1 \in \real^2$, and some number $\alpha \geq 0$, minimize the functional 
\beq\label{goalfun}
J(x)\, = \,  \int_{t_0}^{t_1} |\dot x(t)|\, dt + \alpha \int_{t_0}^{t_1} {\langle \dot x(t), v(x(t)) \rangle^2 \over |\dot x(t)|}\ dt
\eeq
over the set $\X$ of all twice continuously differentiable planar curves $x \equiv \{x(t), t \in [t_0,t_1]\}$ with given end\-points $x(t_0)=\tau_0$ and $x(t_1)=\tau_1$ such that $x(t) \in \Xi$ and $|\dx(t)|>0$ for every $t$.} 

In this setting, $\X$ comprises the possible candidates for the actual percept. The first term in (\ref{goalfun}),
$\int_{t_0}^{t_1} |\dot x(t)|\, dt = \int_{t_0}^{t_1} |dx(t)|$,
represents the length of $x$. (The superscript dot denotes the derivative w.\,r.\,t.~the parameter $t$.) The second term accounts for the context--target interaction: its integrand is essentially the square of the cosine of the angle subtended by $v$ and the curve $x$ at the point $x(t),$ hence it measures the deflection from orthogonality along the curve.  The division by $|\dot x(t)|$ is to make the right-hand side of (\ref{goalfun}) invariant under reparameterizations of the ``time" parameter $t$,\footnote{A functional of the form $y \mapsto \int_{t_0}^{t_1} F(y(t),\dot y(t))\, dt$ is invariant under reparameterization if the integrand is {\em 1-homogeneous} in the sense that $F(u,cv) = c\, F(u,v)$ for all $c>0$ and arguments $u,v$ \cite{Dac04}.} so that it depends only on the {\em trace} of the curve $x$, and not on its parameterization. By its coordinate-free formulation, the problem is also invariant under translations and rotations. The number $\alpha\ge 0$, finally, accounts for the strength of the illusory effect. Obviously, for $\alpha=0$ only the length term is being minimized, and the solution of the problem reduces to the straight line between $\tau_0$ and $\tau_1$, that is, to $\tau$. Since the actual percept deviates only slightly from the straight line target, one may anticipate that $\alpha$ should be small.

\begin{rem} \label{Fermat}
{\em 
The above minimum principle is distantly related to Fermat's principle, which characterizes the path of a light ray through an inhomogeneous medium. Indeed, on rewriting the functional (\ref{goalfun}) in the form $x \mapsto \int_{t_0}^{t_1} F(x(t),\dx(t))\, dt$ with integrand
\beqa
F(x(t),\dx(t)) \ceq |\dot x(t)| + \alpha \langle \dot x(t), v(x(t)) \rangle^2 / |\dot x(t)| \nonumber\\ \ceq |\dot x(t)| \left(1 + \alpha \frac{\langle \dot x(t), v(x(t)) \rangle^2}{|\dot x(t)|^2}\right), \label{Fdef}
\eeqa
one sees that the variational problem amounts to minimizing the functional $ x \mapsto \int_{t_0}^{t_1} n(t)\, |d x(t)| $ where 
$$
n(t) = 1 + \alpha\, \langle \dot x(t), v(x(t)) \rangle^2/|\dot x(t)|^2 = 1 + \alpha \cos^2 \angle (\dx(t),v(x(t)))
$$
is the ``refraction index''---which in our case depends not only on the ``medium'' (here: the context) as traversed by the path, via $v(x(t))$, but also on the tangents to the path, $\dx(t)$.}
\end{rem}

\section{Analysis of the variational problem}\label{vpana}

Let us first introduce some notation. With any curve $x \in \X$ we associate two more curves $\rho \equiv \rho_x,\, \rho^\bot \equiv \rho_x^\bot$, called the {\em Frenet 2-gon:} for every $t$, $\rho(t) = \dx(t)/|\dx(t)|$ denotes the tangent direction vector; $\rho^\bot(t)$ denotes the unit (normal) vector obtained when rotating $\rho(t)$ counterclockwise by $90^\circ$, making $\{\rho(t),\, \rho^\bot(t)\}$ a positively oriented orthonormal basis of $\real^2$. Concerning the context, we write $v'(\xi)$ for the total derivative of $v$ at the point $\xi \in \Xi$, which is a linear mapping from $\real^2$ into itself; $v'(\xi)^\ast$ denotes its adjoint. In standard coordinates, $v'(\xi)$ is given by the $2 \times 2$ matrix of partial derivatives of $v$ at $\xi$ (``Jacobian''), with entries $\pd_j v_k(\xi),\, j,k \in \{1,2\}$, where quite generally, $\pd_j$ stands for the partial derivative \wrt\ the $j$-th argument. Finally, 
$$
\omega(\xi) = \pd_1 v_2(\xi) - \pd_2 v_1(\xi)
$$
denotes the {\em rotation} of $v$ at the point $\xi$. For simplicity, we henceforth assume $\Xi = \real^2$. 

\subsection{Euler-Lagrange equation}\label{eueq}

We apply the apparatus of the calculus of variations \cite{Dac04}. Generally, a curve $x \in \X$ at which a functional of the form $x \mapsto \int_{t_0}^{t_1} F(x(t),\dx(t))\, dt$ attains a minimum necessarily satisfies the {\em Euler-Lagrange equation}
\beq \label{eeg}
\frac{d}{dt}\, \nabla_\dx F(x(t),\dx(t)) - \nabla_x F(x(t),\dx(t)) = 0 \qquad \mbox{(for all $t$).}
\eeq
Here $\nabla_x F,\, \nabla_\dx F$ denote the partial gradients of $F$ with respect to the (vector) arguments $x, \, \dx$, respectively. In our special case where $F$ is given by (\ref{Fdef}), the Euler-Lagrange equation becomes 
$$
\left(1\!-\!\alpha \langle \rho,v(x)\rangle^2\right) \dot\rho = -2\alpha\bigg[\! \left[ v(x)\! -\! \langle \rho,v(x)\rangle \rho \right]\frac{d}{dt} \langle\rho,v(x)\rangle  +  \langle \rho,v(x)\rangle \left(v'(x)\! -\! v'(x)^\ast\right)\dx \bigg]\! ,
$$
where for compactness of notation we omitted the parameter $t$. This system of two nonlinear, second-order differential equations reduces in fact to one single equation that concerns the ``normal'' component orthogonal to the solution curve.

\begin{prop} \label{p: ee1}
For $\alpha < 1$ the normal component of the Euler-Lagrange equation is given by
\beq \label{ee1}
\langle \dot \rho, \rho^\bot \rangle = -2\alpha |\dot x| \, \frac{ \langle v(x), \rho^\bot \rangle \langle \rho, v'(x) \rho \rangle + \langle v(x), \rho \rangle \, \omega(x) }{1 - \alpha\, \langle v(x), \rho\rangle^2 + 2\alpha\, \langle v(x), \rho^\bot\rangle^2}\, .
\eeq
\end{prop}  

In the simplest special case of a constant vector field one would expect that the straight line $\tau$ should result as the unique solution to (\ref{ee1}). Indeed, since $v' = 0$ in this case, the right-hand side of (\ref{ee1}) vanishes, which implies $\langle \dot \rho, \rho^\bot \rangle = 0$, hence $\dot \rho=0$, meaning that the direction vector $\rho$ does not change along $x$. Consequently, $x$ is a straight line, and since its endpoints are fixed at those of $\tau$ it follows that $x = \tau$.

\subsection{Connection with Riemannian geometry}\label{rmgeo}

The very formulation of the variational problem \VP\ and its resemblance to Fermat's principle (cf.~Remark \ref{Fermat}) suggest to look for a strictly geometrical interpretation. Such an interpretation can indeed be given for a slight modification of \VP.

Intuitively, the medium, here represented by the context, perturbs the flat Euclidean geometry so that the shortest path between two points is curved rather than straight. Mathematically, such a non-Euclidean {\em Riemannian geometry} \cite{Lau65} requires specifying a metric tensor $G$ on some differentiable manifold by means of which the length of a parameterized curve $x(t),\, t_0\leq t \leq t_1$ in the manifold is, invariantly under reparameterization, expressed as 
\beq\label{length}
L_G(x) = \int_{t_0}^{t_1} \sqrt{\langle \dx(t), G(x(t))\, \dx(t) \rangle}\, dt.
\eeq
The metric tensor attaches to each point $\xi$ of the manifold a positive definite symmetric matrix 
$$G = G(\xi) = \left( g_{j,k}(\xi) \right)_{j,k=1,2}$$
that depends smoothly on $\xi$. The usual Euclidean geometry corresponds to the special case $G = I$, the $2 \times 2$ identity matrix. A curve $z$ is (a segment of) a {\em geodesic} (in the given geometry) if for any two $t_0 < t_1$ the functional $x \to L_G(x)$ in (\ref{length}) attains its minimum among all smooth curves $x$ with the same endpoints $z(t_0),\, z(t_1)$ as $z$ at the curve $x=z$.

Here, the manifold will be identified with the drawing plane $\real^2$.
The metric is, for given $\alpha \ge 0$ and vector field $v$, defined by
\beq\label{metric}
G \equiv G(\xi) = I + 2 \alpha\, v(\xi)\!\otimes\! v(\xi)\qquad (\xi \in \real^2),
\eeq
with entries $g_{jj} = 1+2\alpha\, v_j(\xi)^2,\, g_{jk} = 2\alpha\, v_j(\xi)\, v_k(\xi)\ (j \neq k)$. The rationale for this choice is straightforward: the root of the quadratic form $\langle \dx(t), G(x(t))\, \dx(t) \rangle$ approximates the function $F$ from (\ref{Fdef}) to the first order in $\alpha$. It may thus be expected that minimization of the criteria (\ref{goalfun}) and (\ref{length}) should yield similar solutions. Precise statements are given in the next subsection. 

Hereafter we will refer to the problem of minimizing the functional $L_G: \X \to \real$ as \GP. In stating the following necessary condition, and further below, we shall use subscripts $\alpha$ whenever we want to emphasize that some quantity depends on the parameter $\alpha$ figuring in \GP\ (or \VP).

\begin{prop} \label{p: rg}
Let $\alpha \ge 0$. A curve $\gamma_\alpha \in \X$ that is a solution to \GP\ (i.e., a geodesic) satisfies the (Euler-Lagrange) equation
\beq\label{eerg}
\ddot x = - 2\alpha\, |\dx|^2 \, \left( \Ft_\alpha(x)\, \rho + \Fn_\alpha(x)\, \rho^\bot \right),
\eeq
where for general $x \in \X$
\beqa
\Ft_\alpha(x) \ceq \frac{1}{1+2\alpha}\, \left( \langle v(x),\rho \rangle \langle \rho,v'(x) \rho\rangle  - 2\alpha\,  \langle v(x),\rho\rangle^2\, \langle v(x),\rho^\bot \rangle\, \omega(x) \right), \label{tcomp}\\ \Fn_\alpha(x) \ceq  \frac{1}{1+2\alpha}\, \left( \langle v(x), \rho^\bot \rangle \langle \rho, v'(x) \rho \rangle + \langle v(x), \rho \rangle \, \omega(x)\ \{ 1+ 2\alpha\,\langle v(x),\rho\rangle^2\, \}\right). \label{ocomp}
\eeqa
\end{prop}

The system (\ref{eerg}) is split into its tangential and normal components by forming inner products with $\rho$ and $\rho^\bot$, respectively. In the latter case this gives after division by $|\dx|$ (and noting that $\langle \ddot x/|\dx|, \rho^\bot \rangle = \langle \dot \rho, \rho^\bot \rangle$) the equation
$$
\langle \dot \rho, \rho^\bot \rangle = - 2\alpha\, |\dx| \,\Fn_\alpha(x),
$$
which may be compared to (\ref{ee1}): for small $\alpha$, the right-hand sides of the two equations are almost the same.

\subsection{Approximate shape of the perceptual distortion}\label{asymp}

The purpose of this subsection is to derive approximations to the geodesic $\gamma_\alpha$ which are then used to define what we call the {\em shape} of the perceptual distortion. That shape, denoted $\sigma$, is uniquely given by the target $\tau$ and the vector field $v$; in particular, knowing the parameter $\alpha$ is not required for determining $\sigma$. Using the shape as the fundamental link, we then clarify the connection between the variational problems \VP\ and \GP.

An explicit expression for the geodesic $\gamma_\alpha$ (solution to (\ref{eerg})) generally is not available; however, for small $\alpha$ it can be approximated by means of a rapidly converging iterative procedure. Given $\alpha\ge 0$ and $x \in \X$ we set
\beq\label{Sdef}
S_\alpha(x, \dx) = -2\,|\dx|^2 \, \left( \Ft_\alpha(x)\, \rho + \Fn_\alpha(x)\, \rho^\bot \right),
\eeq
which considered as a function of $t \in [t_0,t_1]$ represents a curve in $\real^2$. The following is essentially the Picard-Lindel\"of scheme for the iterative solution of an ordinary differential equation (system). Let a sequence of curves $x_{\alpha,n} \in \X$ be defined as follows. One starts with $x_{\alpha,0}= \tau$, the target line, which we take to be parameterized as 
$\tau(t) = \tau_0 + T^{-1}\, (t-t_0)\left(\tau_1 - \tau_0\right),\, T = t_1-t_0$; for $\, n = 1,2,\ldots$,
\beqa\label{it1}
\dx_{\alpha,n+1}(t) \ceq b_{\alpha,n} + \alpha \int_{t_0}^t S_\alpha(x_{\alpha,n}, \dx_{\alpha,n})(u)\, du,\\ 
x_{\alpha,n+1}(t) \ceq \tau_0 + \int_{t_0}^t \dx_{\alpha,n+1}(u)\, du.\label{it0}
\eeqa
The side condition $x_{\alpha,n+1}(t_1) = \tau_1$, {\em i.e.} $x_{\alpha,n+1} \in \X$, is achieved by putting 
\beq\label{bndef}
b_{\alpha,n} = T^{-1} \left( \tau_1 - \tau_0 - \alpha \int_{t_0}^{t_1} \int_{t_0}^t S_\alpha(x_{\alpha,n}, \dx_{\alpha,n})(u)\, du\, dt \right).
\eeq

\begin{prop} \label{p: picard}
Suppose that the mapping $\xi \mapsto v(\xi)$ is twice continuously differentiable in a neighborhood of the target $\tau$. Then there is $\alpha^* \in (0,1)$ and a constant $C$ such that the following holds: for every $0\le \alpha \le \alpha^*$ there exists a solution $\gamma_\alpha \in \X$ to eq.~(\ref{eerg}) such that
\beq\label{expconv}
||x_{\alpha,n} - \gamma_\alpha||_\infty = \sup\nolimits_{\, t_0 \le t \le t_1}\, |x_{\alpha,n}(t) - \gamma_\alpha(t)| \le C \alpha^{n+1} \quad (n=0,1,2,\ldots).
\eeq
\end{prop}

This means that for each sufficiently small $\alpha$ the sequence $x_{\alpha,n}$ converges exponentially fast to a geodesic, $\gamma_\alpha $. By (\ref{expconv}), already the first iteration, $x_{\alpha,1}$, equals $\gamma_\alpha$ up to terms of order $O(\alpha^2)$, which will prove sufficiently accurate for our purposes. On the other hand, $\gamma_\alpha$ (or $x_{\alpha,1}$) differs from $\tau$ by terms of order $O(\alpha)$, which suggests an {\em ansatz} $\gamma_\alpha \doteq \tau + \alpha \sigma$ wherein $\sigma$ would represent the limit as $\alpha \to 0$ of the rescaled deflection $(\gamma_\alpha - \tau)/\alpha$ of $\gamma_\alpha$ from $\tau$.\footnote{%
Thus far, $\alpha$ was a fixed parameter, assumed ``small.'' In the following we conceive of $\alpha$ as an `order parameter' indexing a {\em family} of problems \VP$_\alpha$, \GP$_\alpha$, to be studied asymptotically as $\alpha \to 0$.}
As such, $\sigma$ describes the approximative {\em shape} of this deflection. The scheme (\ref{it1}) to (\ref{bndef}) suggests that $\sigma$ should be given by the conditions $\ddot \sigma = S_0(\tau,\dot \tau)$ and $\sigma(t_0)=\sigma(t_1) = 0$ via a twofold integration, 
\beq\label{sgchar}
\sigma(t) = \int_{t_0}^t \int_{t_0}^s\, S_0(\tau,\dot \tau)(r)\, dr\, ds - T^{-1}(t-t_0) \int_{t_0}^{t_1} \int_{t_0}^s\, S_0(\tau,\dot \tau)(r)\, dr\, ds.
\eeq
For a more explicit description, note first that the 2-gon for the straight line $\tau$ is constant along $\tau$. We denote the corresponding pair of orthonormal vectors as $\rho_0,\, \rho_0^\bot$; thus $\rho_0 = (\tau_1 - \tau_0)/\ell$ with $\ell = |\tau_1 - \tau_0|$ the length of $\tau$, and $\dot \tau = T^{-1}\ell \rho_0$. Observing (\ref{Sdef}), (\ref{tcomp}) and (\ref{ocomp}) one then finds that
\beqa
\ddot \sigma = S_0(\tau,\dot \tau) \ceq -2\, (\ell/T)^2 \left(\Ft_0(\tau)\, \rho_0 + \Fn_0(\tau)\, \rho_0^\bot \right) \label{sgexpl}\\ 
\ceq -2\, (\ell/T)^2 \bigg( \big[\langle v(\tau),\rho_0 \rangle \langle \rho_0,v'(\tau) \rho_0\rangle\big]\, \rho_0 
\nonumber\\ \!\!\! && \qquad\qquad\!\!
+\, \big[\, \langle v(\tau), \rho_0^\bot \rangle \langle \rho_0, v'(\tau) \rho_0 \rangle + \langle v(\tau), \rho_0 \rangle \, \omega(\tau) \big]\, \rho_0^\bot \bigg). \nonumber
\eeqa
{\em The approximation $\widehat x_\alpha= \tau + \alpha \sigma$ will represent our final guess (``prediction'') for the (biased) percept of the target.} Let us say that a certain curve $\eta$ is the approximate shape of the deflections of a family of curves $y_\alpha \in \X\, (\alpha >0)$ from the target, or briefly, {\em the shape (of $y_\alpha$),} if $||y_\alpha - \tau - \alpha \eta||_\infty = O(\alpha^2)$ as $\alpha \to 0$. For example, the shape of $\widehat x_\alpha$ is $\sigma$ (trivially).

\begin{prop} \label{p: appsoln}
Under the conditions of Proposition \ref{p: picard} the following holds.
\beq\label{gapp}
||\gamma_\alpha - \widehat x_\alpha||_\infty = O(\alpha^2) \qquad (\alpha \to 0).
\eeq
Moreover, curves $y_\alpha \in \X\, (\alpha >0)$ with shape $\eta $ satisfy the Euler-Lagrange equation (\ref{ee1}) up to terms of the order $O(\alpha^2)$ as $\alpha \to 0$\footnote{%
The observant reader will notice that we consider approximations at two different levels: the level of solution curves in case of problem \GP\ (first statement), and the level of Euler-Lagrange equations in case of problem \VP\ (second statement). The latter transition frees us from having to refer to `solutions to eq.~(\ref{ee1})' the existence of which is unclear in case of problem \VP\ (other than with \GP).}
if and only if $\langle \eta, \rho_0^\bot \rangle = \langle \sigma, \rho_0^\bot \rangle$. 
\end{prop}

The first statement implies that the geodesics $\gamma_\alpha$ as well as the approximations $x_{\alpha,n}, \, n \ge 1$ all share the same shape as $\widehat x_\alpha$, namely $\sigma$. (This follows from (\ref{gapp}) and (\ref{expconv}), which together give $||x_{\alpha,n}- \widehat x_\alpha||_\infty = O(\alpha^2)$.) In particular, $\widehat x_\alpha$ approximates the solution $\gamma_\alpha$ to eq.~(\ref{eerg}) to the first order in $\alpha$ (i.e., up to terms of order $O(\alpha^2)$). On the other hand, $\widehat x_\alpha$ also represents a first-order approximate solution to eq.~(\ref{ee1}) since it trivially satisfies the if-condition in the second statement. In fact, {\em any} first-order approximate solution to (\ref{ee1}) necessarily has, to first order, the same lateral deflection from the target as $\widehat x_\alpha$, in that the normal components of their respective shapes are identical. 

The important conclusion here is that the {\em same} curve, $\widehat x_\alpha$, represents an approximate solution, accurate up to terms of order $O(\alpha^2)$, to {\em both} problems \VP\ and \GP\ simultaneously. Hence, the phenomenologically motivated and the geometrical approaches leading to the variational problems \VP\ and \GP, respectively, yield, to first order, identical predictions for the shape of the perceptual distortion. In that sense, the two approaches are equivalent.

\begin{rem} \label{sign}
{\em 
Since $\langle \ddot {\widehat x}_\alpha, \rho_0^\bot \rangle/\alpha = \langle \ddot \sigma, \rho_0^\bot \rangle = -2(\ell/T)^2\, \Fn_0(\tau)$, by (\ref{sgexpl}), the sign of
$$
\Fn_0(\tau) = \langle v(\tau), \rho_0^\bot \rangle \langle \rho_0, v'(\tau) \rho_0 \rangle + \langle v(\tau), \rho_0 \rangle \, \omega(\tau)
$$
determines whether $\widehat x_\alpha$, when traveled through from $\tau_0$ to $\tau_1$, is bending to the left-hand side (sign $\!\Fn_0(\tau) = -1$) or to the right-hand side (sign $\!\Fn_0(\tau) = +1$), respectively. Therefore, the qualitative winding behavior of $\widehat x_\alpha$ can be read off already from that sign; knowing the shape $\sigma$ {\em completely} is not required for this purpose.} 
\end{rem}

Conveniently, $\sigma$ depends only on the known quantities $v(\tau)$ and $\tau$, making $\widehat x_\alpha$ easily calculable for any trial parameter $\alpha$. This allows for a straightforward implementation of the method of compensatory measurement in our experimental study described in Section \ref{exp}. 

\section{Examples}\label{exs}

Here we introduce three families of context curves forming the streamlines of the vector field $v$. We represent such a family by means of a real-valued smooth function $c(u,\theta)$ depending on two real arguments $u, \theta$ such that $c$ is strictly increasing in $\theta$ for each fixed $u$. The context curves $u \mapsto C_\theta(u) = (u,c(u,\theta))$ then do not intersect for different $\theta$s, and we may assume that for every point $\xi=(\xi_1,\xi_2)$ in some region $\Xi \subseteq \real^2$ there exists $\theta = \vartheta(\xi_1,\xi_2)$ such that $c(\xi_1,\vartheta(\xi_1,\xi_2)) = \xi_2$.\footnote{For notational convenience coordinate vectors are written as row vectors.} Within this setting, one can calculate the crucial quantities $\Ft_\alpha,\, \Fn_\alpha$ explicitly in terms of partial derivatives of $c$.

Of particular interest is the behavior of the normal and tangential components, $\Fn_\alpha$ and $\Ft_\alpha$, along the target. Suppose that $\tau$ is the horizontal line segment between $\tau_0 = (-\ell/2,0)$ and $\tau_1 = (\ell/2,0)\ (\ell>0)$, parameterized by $t \equiv u \equiv \xi_1 \in [-\ell/2,\ell/2]$. Then for $\alpha=0$ we have
\beqa \label{n0xpl}
\Fn_0(\tau)\, \equiv\, \Fn_0\, \ceq \, \left[\, \pd_{11}^2 c \left(1 - (\pd_1 c)^2\right) \, + \, \pd_{12}^2 c \cdot (\pd_1 c)^3/\pd_2 c\, \right] \big/ \left[\,1+ (\pd_1 c)^2\, \right]^2, \\
\Ft_0(\tau)\, \equiv\, \Ft_0\, \ceq \, -\pd_1 c \left(\pd_{11}^2 c -  \pd_{12}^2 c \cdot \pd_1 c/\pd_2 c\right) \big/ \left[\,1+ (\pd_1 c)^2\, \right]^2,\label{t0xpl}
\eeqa
wherein the partial derivatives  are evaluated at the arguments $(\xi_1,\vartheta(\xi_1,0))$. (Note that one has $c(\xi_1,\vartheta(\xi_1,0))=0$ by the definition of $\vartheta$, so the parameter $\theta$ for which the curve $C_\theta$ crosses the target at the point $(\xi_1,0)$ is $\theta = \vartheta(\xi_1,0)$.) We only give the expressions for the primarily important quantity $\Fn_0$, which describes the lateral deflection of the percept from the target. 

The following three types of functions $c$ will be considered.

\medskip\noindent
{\bf Type 1} (vertical shifts): 
$c(u,\theta) = q(u) + \theta$ for some given function $q$. Then
$$
\pd_1 c = q', \quad \pd_2 c = 1, \quad \pd_{11}^2 c = q'', \quad \pd_{12}^2 c = 0
$$
(primes here denote derivatives \wrt\ $u$), whence
$$
\Fn_0 \, = \, q'' \left(1 - q'^{\, 2}\right) \big/ \left(1 + q'^{\, 2} \right)^2.
$$
If $q$ is even and convex then $\Fn_0 > 0$ at least in the central part of $\tau$, since $q'(0)=0$. Thus in view of Remark \ref{sign}, our principle predicts that the curve appears concave there (bending downward away from the origin. This fits with the perceived curvatures in Figs.~3a, 3b, 3d, as well as Fig.~1d (lower edge) or Fig.~1c (upper line). Of course, by symmetry the converse holds if $q$ is concave instead of convex;  see Figs.~3c, 3e, 3f, 1d (upper edge), 1c (lower line).

\medskip\noindent
{\bf Type 2} (dilation): 
$c$ is of the form $c(u,\theta) = \theta q(u) - a$ with a constant $a>0$ and a function $q$ satisfying $q(u) > q(0) > 0,\, 0<|u| \le \ell$. Here
$$
\pd_1 c = \theta q'\, , \quad \pd_2 c = q, \quad \pd_{11}^2 c = \theta q'', \quad \pd_{12}^2 c = q'
$$
with $\theta = \vartheta(\xi_1,\xi_2) = (a + \xi_2)/q(\xi_1)$ (and $\xi_1=u$). Hence along $\tau$, where $\xi_2=0$ or $\theta = a/q$,
$$
\Fn_0 \, =  \left[\, \left(1 - (a q'/q)^2\right) aq''/q \, + \, a^3\, (q'/q)^4\, \right] \big/ \left[\, 1 + (a q'/q)^2\, \right]^2.
$$
Again, $\Fn_0$ is positive (negative) near the origin if $q'(0)=0$ and $q$ is convex (concave) thereabouts. The conclusion in regard to the perceived curvature thus is the same as for type 1.

\medskip\noindent
{\bf Type 3} (segments of concentric circles): 
$c(u,\theta) = \sqrt{\theta^2-u^2} - a,\ |u| \le \theta$ where $a$ is a positive constant and $\theta > a$. The curves $C_\theta$ represent concentric upper half circles intersecting the $x$-axis at the points $\pm \sqrt{\theta^2-a^2}$. Observing $\theta = \vartheta(\xi_1,\xi_2) = \sqrt{\xi_1^2+(a+\xi_2)^2}$ one finds that for $\xi_2=0$ the numerator of $\Fn_0$ in (\ref{n0xpl}) equals the constant $-1/a$, and
$$
\Fn_0 \, = \, -a^{-1} \left(1 + (\xi_1/a)^2\right)^{-2}.
$$
In particular, $\Fn_0 <0$, and the principle predicts that the perceived curve should be convex (bending upward away from the origin), in agreement with, e.g., Figs.~1d, 1f, upper edge. 

\medskip
It should be noted that the conclusions regarding curvature are similar for the various context types. This suggests a simple rule of thumb: the percept tends to be bent in the opposite direction as the context curves. See e.g.~Fig.~2a, 2b, where the curvatures of the pattern and the shape $\sigma$ differ between the center and the margins, in opposite ways.

\begin{figure}[t]
\begin{tabular}{ll}
a) & b)\\
\insfig{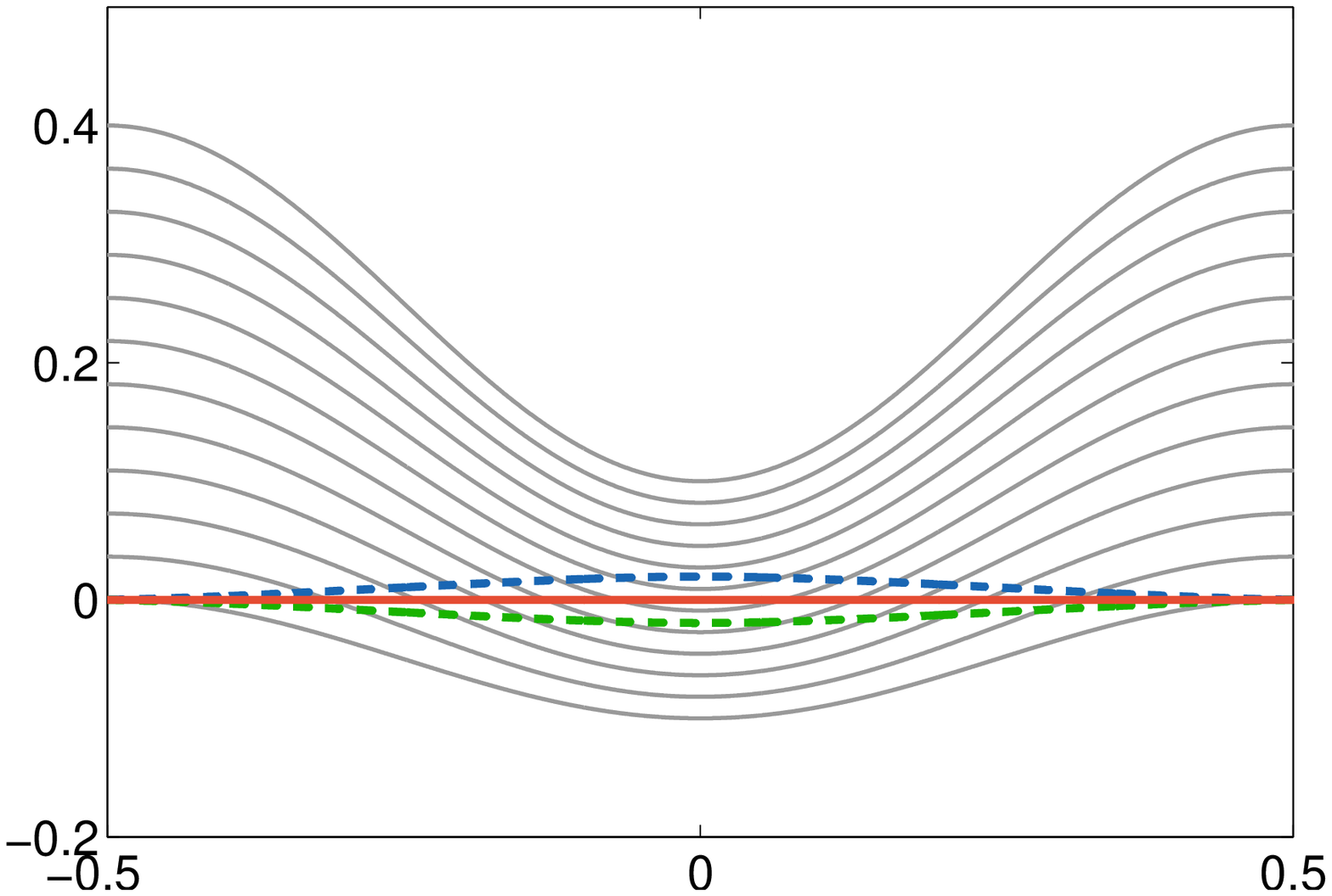}{0.375} & \insfig{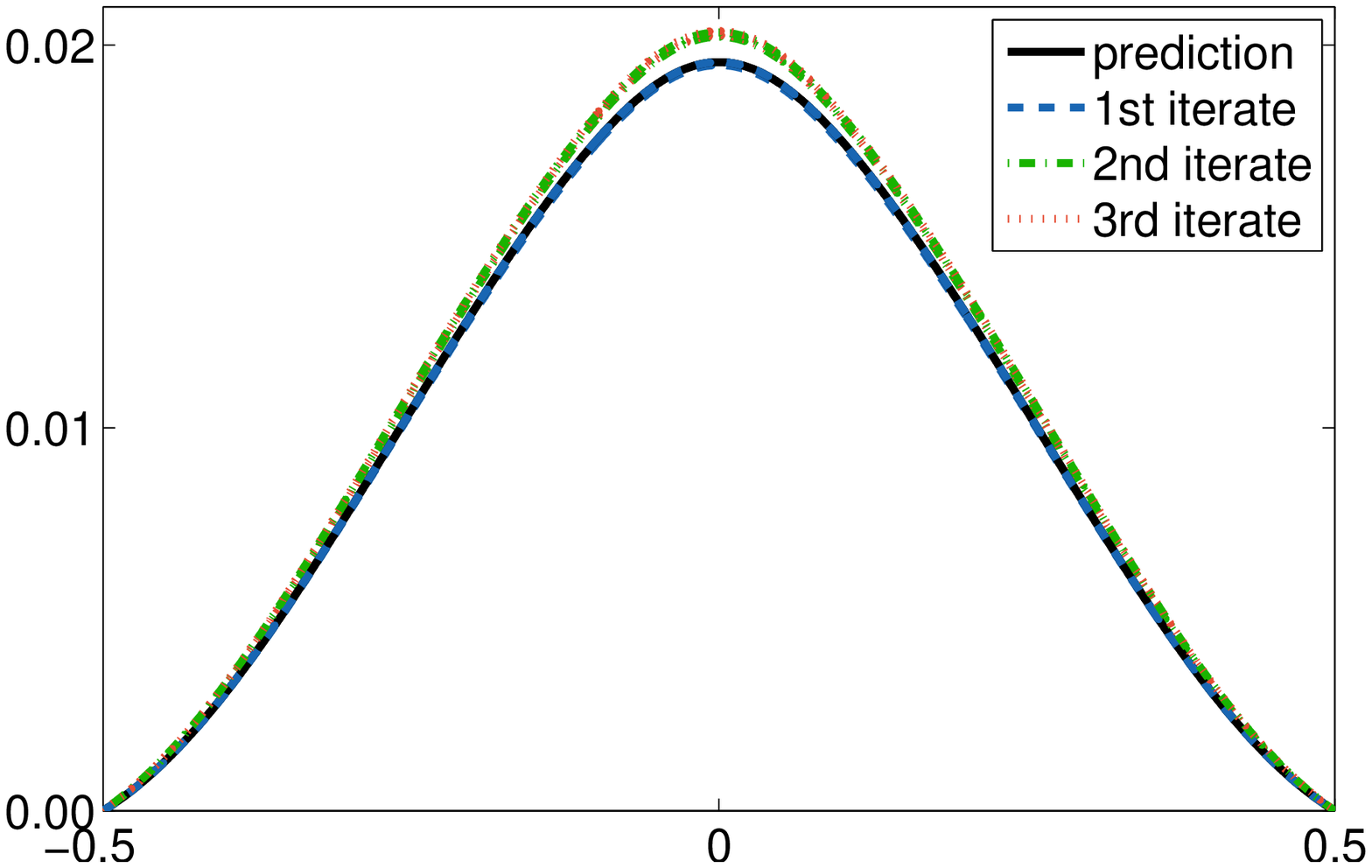}{0.375}
\end{tabular}

\medskip
\fns
Figure~2:
a) Illustration of context-induced effects: Straight line target $\tau$ (red line) embedded in a type~2 context (array of curves). The dashed line above $\tau$ represents our prediction for the average observer's percept, the lower (dashed) line is counterbalanced so as to be perceived as straight by the average observer. See text for detailed explanation.\ ---
b) Comparison of various approximations. Shown are: (i) the prediction $\widehat x_{\bar\alpha}$, and the first three iterations converging to the geodesic $\gamma_{\bar\alpha}$; (ii) the analogous curves when the tangential component is ignored. See text. Note that up to scaling all curves share approximately the same form, namely that of the shape of the perceptual distortion, $\sigma$.
\end{figure}

Let us discuss this example in more detail. The context in Fig.~2a is of type 2, with $q(u) = 1 + \sin^2(\pi u),\, |u| \le 1/2$, $a = 0.239$, and twelve $\theta$'s equally spaced between 0.1 and 0.3. The two lines above and below the target $\tau$ are determined as $\widehat x_\pm = \tau \pm \bar\alpha \sigma$, respectively, where $\sigma$ is the shape computed numerically via (\ref{sgchar}), (\ref{sgexpl}) using (\ref{n0xpl}), (\ref{t0xpl}); and $\bar\alpha = 0.05$ is the average $\alpha$ value (across trials and participants) obtained in the experiment described in Section \ref{exp}. Note that the target appears slightly bent upward in the middle, and this effect is roughly doubled when $\widehat x_+$, which is our prediction for the average observer's percept, is drawn within the same context. Conversely, subtracting the distortion as in $\widehat x_-$ removes the perceived curvature for the average observer.\footnote{%
In the experiment, only {\em one} line was shown at a time (together with the context). The ensuing perceptual bias may well be different from the one seen in Fig.~2a; and of course, it may differ between observers.}
This `compensation principle' is used in the implementation of our experiment.

Fig.~2b shows the prediction $\widehat x_+$ along with the first three iterates $x_{\bar\alpha,n},\, n=1,2,3$ computed via (\ref{Sdef}) to (\ref{bndef}), which approximate the exact geodesic $\gamma_{\bar\alpha}$. Also plotted are four curves obtained in the very same way except that the term $S_{\bar\alpha}(x,\dx)$ throughout is replaced by the term $-2|\dx|^2\, \Fn_0(x) \rho^\bot$ ignoring the tangential component. These eight curves come in two groups of four curves each which within groups are almost identical. The lower quadruple consists of $\widehat x_+,\ x_{\bar\alpha,1}$, and their counterparts computed with $-2|\dx|^2\, \Fn_0(x) \rho^\bot$ instead of $S_{\bar\alpha}(x,\dx)$; the upper quadruple consists of the respective second and third iterates. Similar results were found for all cases considered.

\subsection{Gaussian curvature}\label{Gaucurv}

In Section \ref{rmgeo}, we attached to each context a metric tensor $G$ via the associated vector field. An intrinsic property of the geometry induced by $G$ is the Gaussian curvature, $K$. This is a scalar quantity that describes how, and how strongly, the corresponding manifold (here $\real^2$) is deformed at each of its points \cite{Lau65}. An approximation to $K$ valid for our setting is
\beq\label{GK}
K = 2\alpha\, \FC + O(\alpha^2), \qquad \mbox{where}\quad  \FC = v_1\, \pd_2 \omega - v_2\, \pd_1 \omega - \omega^2
\eeq
depends on the rotation $\omega$ of the respective vector field $v$ in the first place. 

For contexts of type 1 and 2, $\FC$ (hence $K$) turns out to vary across the manifold, and to assume positive as well as negative values. The geometry thus does not reduce to one of the classical non-Euclidean geometries (elliptic, hyperbolic, etc.) where $K$ is constant. This is different with contexts of type 3: here $\FC$ can be shown to vanish identically, which implies a flat (essentially Euclidean) geometry. 

\section{Experiment}\label{exp}

Our approach predicts the shape of the perceived distortion of the target as given by the expression $\sigma$ introduced in Section \ref{asymp}. The magnitude of the distortion is determined by the parameter $\alpha$, which has to be estimated empirically. For that purpose we carried out an illustrative experiment, using the method of {\em compensatory measurement}: a line distorted in the opposite direction is presented to the observer, who adjusts $\alpha$ until a straight line is perceived. The rationale for this procedure is clear: If $\tau$ is (approximately) perceived as $\tau+\alpha\sigma$, then for small $\alpha$, $\tau-\alpha\sigma$ will (approximately) be perceived as $\tau$.

The stimuli were constructed for six different contexts, as shown in Fig.~3. Each stimulus consisted of a sequence of 21~Encapsulated Postscript pictures (`frames'), displaying a constant array of context curves, drawn black on a white background, with superimposed curved lines of the form $\tau-\alpha_k\,\sigma\ (k = 1,\ldots,21)$, the $\alpha_k$s being equally spaced in the (sufficiently large) interval $[-.11,.29]$. The target $\tau$ was always a horizontal straight line segment, drawn in red for easier visual identification. The 21 frames belonging to a single stimulus were combined to a single multipage \PDF\ file,\footnote{%
Two \PDF\ files were prepared for each context, with the frames sequence in the `forward' order $\alpha_1,\ldots,\alpha_{21}$, and in the `backward' order $\alpha_{21},\ldots,\alpha_1$. These two versions were used alternately in each experimental session (see below). The case $\alpha = 0$ (exactly straight line) was {\em never} contained in the sequence.}
which was displayed on a \LCD\ monitor watched binocularly from a distance of 100~cm. The observers' task was to scroll through the sequence of frames and to indicate that one where the red line appeared to them as most similar to a straight line. Each trial thus resulted in an estimate $\widehat\alpha$ of the model parameter~$\alpha$.

\begin{figure}[t]		
\insfigc{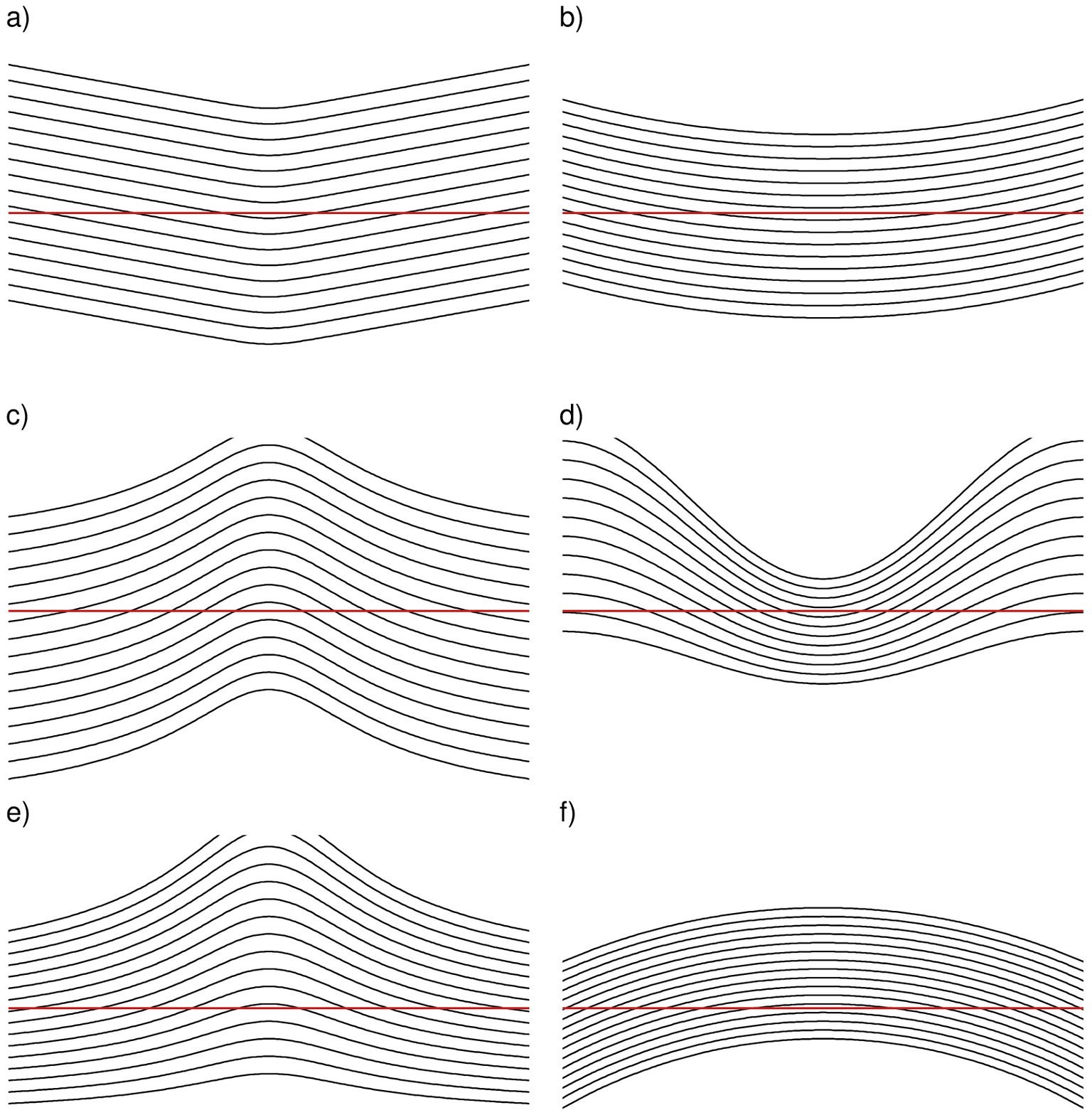}{0.75}

\vspace{-5mm}
\fns
Figure~3:
Six context patterns used in the reported experiment. Classification according to Section~\ref{exs}: a--c): type~1; d,\,e): type~2; f): type~3. Shown are stimuli for $\alpha=0$, \ie, superimposed red lines are exactly straight lines.
\end{figure}

Nine observers participated in the experiment. Each participant was presented stimuli of six different classes (contexts), in a randomized order, and six trials were done with each stimulus class.\footnote{%
Three of the six trials were run with the `forward' and three with the `backward' frames sequence to avoid possible directional bias in the observer's response.}
The study thus yielded a total of 9~(observers) $\times$ 6~(contexts) $\times$ 6~(repetitions) = 324 estimates of~$\alpha$. The complete data set is presented in Fig.~4. All $\alpha$-estimates are positive, in accordance with the predicted direction of the distortion. Despite interindividual differences in susceptibility to the illusory effect, the pattern of the $\alpha\!$s (disregarding the average magnitude) is remarkably similar across subjects. 

For further analysis, we attempted to decompose the responses into an individual factor and a factor depending on the context.\footnote{%
The strength-of-effect parameter $\alpha$ may reflect a variety of factors, including factors that depend on $\sigma$.}
Let $\alpha(i,c)$ denote the average $\alpha$-estimate across trials reported by observer $\# i$ for context $\# c$. If the $\alpha(i,c)$ are proportional to the product of an individual factor, $\eta(i)$, times a context-dependent factor, $\kappa(c)$, then dividing these factors out renders the thus normalized responses $\alpha(i,c)/\left(\eta(i) \kappa(c)\right) \equiv \widetilde \alpha(i,c)$ constant. In that (ideal) case one can argue that those two factors fully ``explain'' the (systematic) variation in the data. The goal thus is to find subject- and context-dependent factors reducing the variation in the $\widetilde \alpha(i,c)$ as far as possible.

\begin{figure}[t]
\insfigc{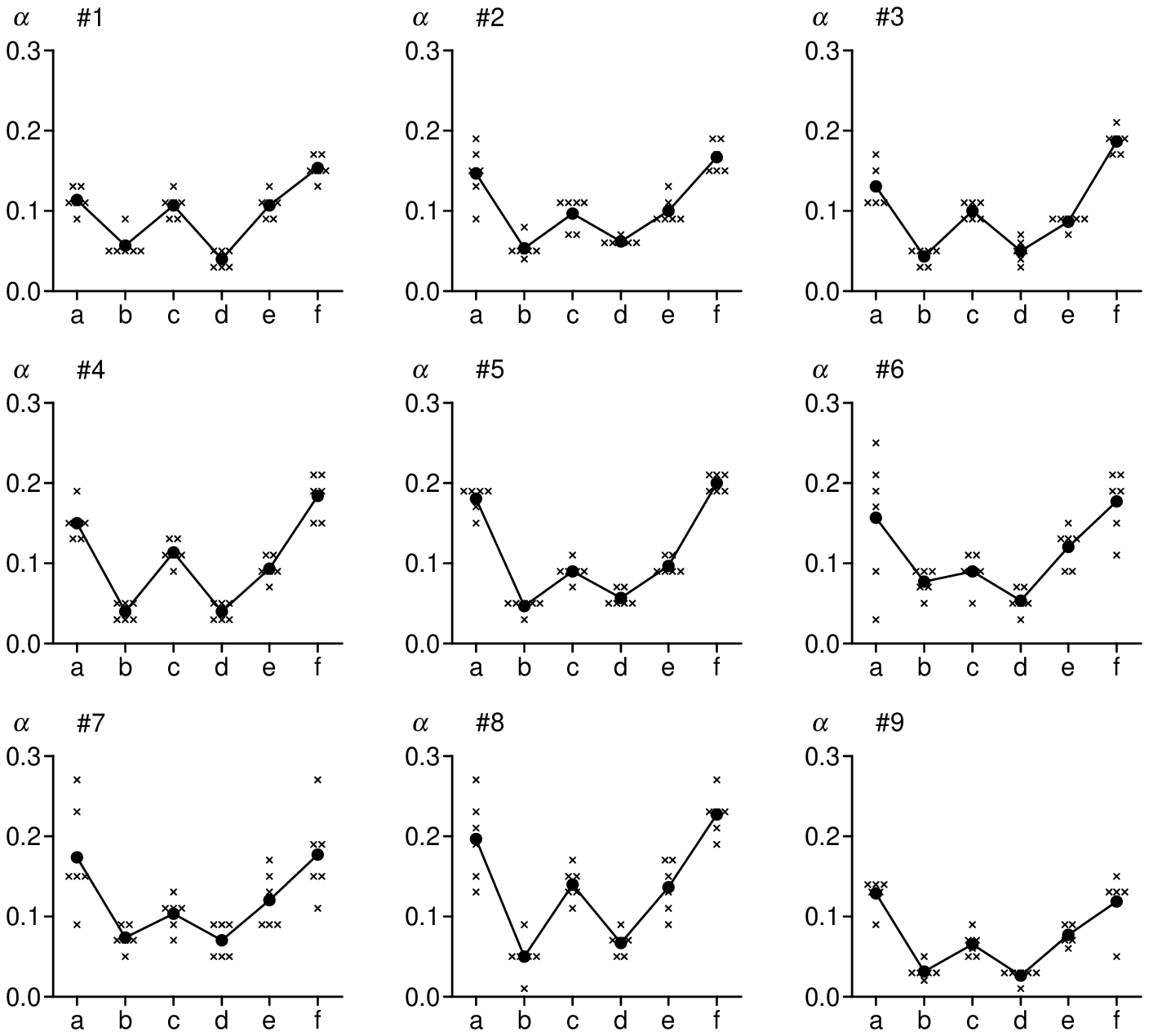}{0.75}

\fns Figure 4: Results of the experiment described in Section~\ref{exp}. Each of the nine panels displays the complete data set from an individual participant  (36~trials). Absciss{\ae}: context patterns labeled as in Figure~2; ordinates: estimates of $\alpha$; cross marks: single-trial estimates; filled circles (connected): arithmetic means.\par
\end{figure}

A natural choice for $\eta(i)$ is the average of the $\alpha(i,c)$ across contexts, $\eta(i) = \overline{\alpha(i,\cdot)}$. Suitable candidates for the factor $\kappa(c)$ could be various geometrical quantities related to, for example, the number and angles of the context-target intersections, or the curvature of the context lines. The best among those considered was\footnote{%
Noteworthily, $\kappa$ depends only on the component of $\sigma$ orthogonal to $\tau$, so that $\kappa^2$ represents a kind of ``energy'' contained in the lateral deflection of the percept from the target.}
$$
\kappa =  \sqrt{\frac{1}{T} \int_{t_0}^{t_1} \langle \dot\sigma(t), \rho_0^\bot \rangle^2\, dt}\, ,
$$
where ``best'' means the following. Let individually normalized response {\em profiles} be defined as $\pi(i,c) = \alpha(i,c)/\eta(i)$, and let $\pi(c) = \overline{\pi(\cdot,c)}$ denote their average across observers (considered as functions of $c$, each). The above $\kappa$ was best in the sense that division by this term maximally reduced the coefficient of variation, namely from CV = 0.48 for the profile $\pi(c)$ to CV = 0.18 for the profile $\pi(c)/\kappa(c)$ additionally normalized by $\kappa(c)$. The individual profiles $\pi(i,c)$ along with the group mean $\pi(c)$ are shown in Fig.~5a. Similarly, Fig.~5b presents the respective profiles additionally normalized by context, $\pi(i,c)/\kappa(c) = \widetilde \alpha(i,c)$ and $\pi(c)/\kappa(c) = \overline{\widetilde \alpha(\cdot,c)}$.

\begin{figure}[t]
\insfigc{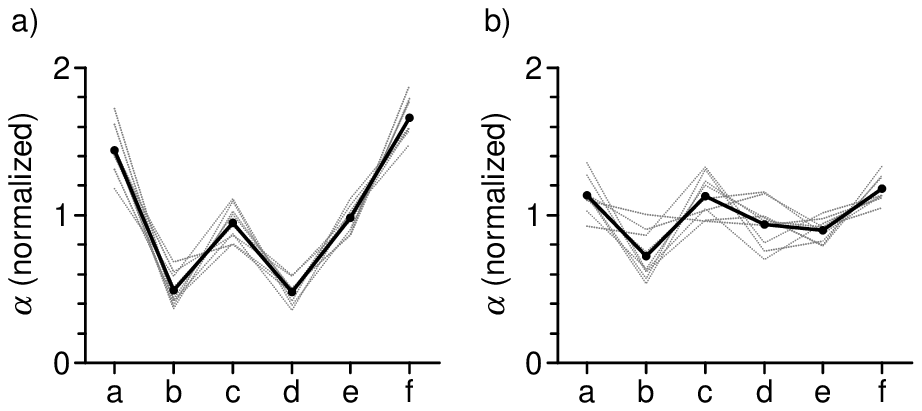}{1.125}

\fns
Figure~5:
Individual (grey, dotted lines) and group (black, solid line) response profiles. a)~First step: responses $\alpha(i,c)$ normalized intra-individually; coefficient of variation = 0.48. b)~Second step: profiles resulting from Step 1 further divided by context-dependent factors $\kappa(c)$; coefficient of variation = 0.18. Details are provided in Section~5.
\end{figure}

\section{Discussion}\label{disc}

The present paper marks but one step in our approach to the study of visual field geometry. Admittedly, the approach presented here has certain limitations. Some of these limitations follow naturally from our decision for a ``phenomenological,'' \ie\ purely descriptive theory of the {\GOI} phenomena \cite{Wac10}, disregarding possibly underlying neurophysiological or neuropsychological mechanisms. Other limitations reflect the momentary state of development of the theory, and will hopefully be overcome at later stages:

1.~Modeling the context by a continuous vector field relies upon a convenient, yet unrealistic idealization; in reality, the context always consists of an array of finitely many distinct curves. To what extent this idealization is justifiable remains an open question.\footnote{%
One might hope that a variable density of the target--context intersection points could be mimicked by admitting nonlinear parameterization of the context curves, \eg, by working with functions $(u,\theta) \mapsto c(u, \phi(\theta))$ where $\phi$ depends nonlinearly on $\theta$. It turns out, however, that the terms $\Fn_0(\tau),\, \Ft_0(\tau)$, hence the shape of the distortion, are invariant under such reparameterizations.}

2.~Optical properties (color, background brightness, figure/background contrast, etc.)\ of the stimulus as well as its global geometric properties (relative size in the visual field, orientation \wrt\ gravicentric coordinates, etc.)\ have no representation in the present approach. One may expect that these properties do not affect the form but only the magnitude of the perceptual distortion, and can thus be accounted for by the ``illusion strength'' parameter $\alpha$. Experimental studies must decide which parameters of the stimulus may enter the model via parameter $\alpha$.

3.~The assumption of local interactions [Section~\ref{VP}, sub~(a)] implies that ``holistic'' properties of the context pattern (symmetry, presence or absence of ``focal'' points, etc.) are plainly ignored. While we feel that the global, ``scenic'' appearance of the context has been over-interpreted in some explanatory approaches, \eg\ \cite{Tau54, Gre63,Day72}, we cannot \emph{a~priori} exclude that such holistic properties may play a modulating r\^ole. These aspects, as well as those mentioned above sub 2, call for more experimental research.

4.~The variational approach with fixed endpoints imposes a severe restriction on admissible percepts. For example, the present framework does not allow to treat the well-known Z\"ollner illusion \cite{Zoe60}, where the target lines are perceived as tilted, but preserve their straight line appearance. Moreover, extension from straight lines to targets of simple geometric forms---\eg\ distortions of circles to oval shapes in Hering-like contexts \cite{Ehr25,Orb39}---is certainly desirable. 
More generally, one may think of targets representing geodesics in some Riemannian basis geometry that is perturbed by the context similarly as the Euclidean metric tensor $I$ is perturbed by the term $2 \alpha\, v\!\otimes\! v$ in (\ref{metric}).

A demarcation of our approach against so-called ``field theories'' of {\GOI}s \cite[pp.~167--170]{CoGi78} appears necessary. In our approach, the vector field is a convenient mathematical representation of the context pattern. By contrast, some researchers think of a vector field induced in the neural substrate by the context part of the stimulus as a {\em physical} entity. This idea, originating in early theories of psychophysical isomorphism \cite{Koe20}, inspired some modeling/explanatory approaches \cite{BrVo37,Orb39,Eri70,Mot70} that remained mostly on a qualitative or semi-quantitative level. 

Closer in spirit to the present approach is the work of Hoffman \cite{Hof66,Hof71} and Smith \cite{Smi78} wherein, too, the``realist'' concept of a (neuro)physical field was abandoned in favor of a purely formal, mathematical treatment. These developments based on vector fields and Lie derivatives represent a line of research parallel to ours: they, too, assume a local interactions and angular expansion hypothesis, and yield a prediction for the perceptual distortion of a form similar to ours. However, neither did these works make use of the calculus of variations for the derivation of the distortion, nor did they establish a connection with Riemannian geometry. 

A Riemannian geometry for visual perception was in fact derived by Zhang and Wu \cite{ZW90} by elegant considerations of perceptual coherency of a visual object under rigid translations. Zhang and Wu build on the image intensity function and properties of motion detectors, and derive an affine connection depending on derivatives of the image function as the fundamental constituent of the geometry. Their concepts are not easily seen to be applicable to the present setting, however, which deals exclusively with static percepts. One difference concerns the Gaussian curvature, $K$, implied by the respective geometries. In \cite{ZW90}, $K \equiv 0$ always, making the geometry flat or pseudo-Euclidean, whereas in our approach $K$ may assume positive and negative values across the manifold, and may also vanish identically, depending on the context; cf.~Section \ref{Gaucurv}. Furthermore, Zhang and Wu's geodesics are ``perceptually straight,'' whereas ours are not, being the curved percept of a ``physically straight'' line.

Summarizing, we believe that our approach, in spite of its limitations discussed above, has its undisputable merits and potential for further developments:

1.~Approximate solutions to the variational problem\footnote{%
Variational calculus is a widely used tool in vision science, computer vision (e.g., \cite{Fos78, AM03}), and elsewhere. Its use in, and specific application to, the current setting seem to be new, however.} 
introduced in Section~\ref{VP} yield phenomenologically correct predictions for the perceptual distortion once the only free parameter of the model, $\alpha$, has been determined (experimentally). This is evidenced by the fact that properly counter-distorted targets appear, in a given context, as straight lines without perceivable residual distortion. A more thorough validation of the predicted shape of the distortion would certainly be desirable, but this is beyond the scope of the present work.

2.~The experimental data reported in Section~\ref{exp} show a remarkable stability of the illusory effect across participants and context types. This finding supports the notion that \GOI s are not mere failures of the visual system, but that they reflect intrinsic principles of the structural organization of visual percepts \cite{Met75}. The method of the reported experiment can be used to study dependence of the ``illusion strength'' parameter $\alpha$ on various properties of the stimulus.

3.~Perhaps the most important feature of the mathematical model is its explicitly geometrical interpretation, which allows us to characterize the percept of a straight line as a geodesic in an appropriate, context-dependent Riemannian geometry (Section~\ref{rmgeo}). The analogy between the theory of a (world-)space metric, dependent on the mass distribution, and a possible theory of visual space metric, dependent on the perceptual content of the visual field, has been noticed by several authors \cite{Wag06,Wes08,WaKa09}. An important contribution here is the work of Zhang and Wu \cite{ZW90} who studied the perceptual coherence of a visual object under rigid motions. Here we demonstrate for the purely static case of geometric-optical illusions how the presence of context elements in the visual field perturbs its (initially Euclidean) geometry, as reflected by the metric tensor (\ref{metric}), and how the illusory distortion \emph{naturally} arises from the perturbed geometry.

\section*{Acknowledgments}

The authors are grateful to Steffen Heinze for his suggestions concerning variational calculus and Riemannian geometry, and for valuable remarks on an earlier draft of the paper. Thanks are also due to the Editor and three reviewers for their constructive criticism and recommendations that helped to improve the paper.

\newpage
\section*{Appendix}

\subsection*{A1. The Euler-Lagrange equations for problems \VP\ and \GP}

{\em Proof of Proposition \ref{p: ee1}. }
Using the notation introduced in Section \ref{vpana} we calculate
\beqas
\nabla_\dx F \ceq \frac{\dx}{|\dx|} +  \frac{2\alpha\langle \dot x, v(x) \rangle}{|\dx|} \,v(x) - \frac{\alpha\langle \dot x, v(x) \rangle^2}{|\dot x|^3}\, \dx \\ \ceq \rho + 2 \alpha \langle \rho, v(x) \rangle\, v(x) - \alpha \langle \rho, v(x) \rangle^2\, \rho,\\
\frac{d}{dt}\, \nabla_\dx F \ceq \left(1\!-\!\alpha \langle \rho,v(x)\rangle^2\right) \dot\rho  -2 \alpha \langle \rho,v(x)\rangle\, \rho\,  \frac{d}{dt} \langle \rho,v(x)\rangle \\ \!\!\! && + \ 2 \alpha \, v(x)\, \frac{d}{dt} \langle \rho,v(x)\rangle + 2 \alpha \langle \rho,v(x)\rangle\, v'(x) \dx \\ \ceq
\left(1\!-\!\alpha \langle \rho,v(x)\rangle^2\right) \dot\rho  + 2 \alpha \langle \rho,v(x)\rangle\, v'(x) \dx + 2 \alpha \left[ v(x) - \langle \rho,v(x)\rangle\, \rho\, \right] \frac{d}{dt} \langle \rho,v(x)\rangle,\\
\nabla_x F \ceq \frac{2\alpha\langle \dot x, v(x) \rangle}{|\dx|} \, v'(x)^\ast \dx = 2 \alpha \langle \rho, v(x) \rangle\, v'(x)^\ast \dx,
\eeqas
so the general Euler-Lagrange equation (\ref{eeg}) assumes the form
\beqa\label{ees}
&& \!\!\! \left(1\!-\!\alpha \langle \rho,v(x)\rangle^2\right) \dot\rho \\ \ceq -2\alpha\left\{ \left[ v(x)\! -\! \langle \rho,v(x)\rangle \rho \right]\frac{d}{dt} \langle\rho,v(x)\rangle  +  \langle \rho,v(x)\rangle \left[v'(x)\! -\! v'(x)^\ast\right]\dx \right\}\! .\nonumber
\eeqa

Initially, (\ref{ees}) is a system of two nonlinear, second-order differential equations. However, both sides of (\ref{ees}) are in fact orthogonal to $\rho$ for all $t \in [t_0,t_1]$, meaning that the tangential component is trivial and only the component orthogonal to it matters. To see this, note that
$\dot \rho  = |\dot x|^{-1}\ddot x - |\dot x|^{-3}\langle \ddot x, \dot x \rangle \dot x$,
whence $\langle \dot \rho, \rho \rangle = 0$; moreover,
$$
\left\langle v(x) - \langle \rho,v(x)\rangle \rho, \rho \right\rangle = 0
$$
as well as 
$$
\left\langle \left(v'(x)-v'(x)^\ast\right) \dot x, \rho \right\rangle \, = \,  |\dot x| \left( \left\langle v'(x)\rho, \rho \right\rangle - \left\langle \rho, v'(x)\rho \right\rangle \right)  = \, 0.
$$
Thus effectively, the system (\ref{ees}) reduces to one equation. Now
\beq \label{drcollin}
\dot \rho  = \langle\dot \rho, \rho^\bot \rangle \rho^\bot
\eeq
(since $\langle \dot \rho, \rho \rangle = 0$), so forming the inner product of (\ref{ees}) with $\rho^\bot$ we get the relevant part of the Euler-Lagrange equation (system), 
\beqa
0 \ceq \langle \dot \rho, \rho^\bot \rangle \left(1\!-\!\alpha \langle \rho,v(x)\rangle^2\right) +  2\alpha\langle v(x), \rho^\bot \rangle \frac{d}{dt} \langle\rho,v(x)\rangle \nonumber\\ \!\!\! && \qquad\quad +\ 2\alpha\langle \rho,v(x)\rangle \left\langle\! \left(v'(x)-v'(x)^\ast\right) \dot x, \rho^\bot\! \right\rangle \nonumber\\ \ceq 
\langle \dot \rho, \rho^\bot \rangle \left( 1 - \alpha\, \langle v(x), \rho\rangle^2 + 2\alpha\, \langle v(x), \rho^\bot\rangle^2 \right) \nonumber\\ \!\!\! && \qquad\quad +\  2 \alpha |\dx|\left(\langle v(x), \rho^\bot \rangle \langle v'(x) \rho, \rho \rangle + \langle v(x), \rho \rangle \left(\pd_1 v_2(x) - \pd_2 v_1(x)\right) \right), \label{eeqn1} 
\eeqa
where the second equality follows via (\ref{drcollin}) from 
$$
\frac{d}{dt} \langle\rho,v(x)\rangle  =  \langle \dot \rho, v(x) \rangle + \langle \rho, v'(x) \dx \rangle  = \langle \dot \rho, \rho^\bot \rangle \langle v(x), \rho^\bot \rangle + |\dx| \langle v'(x) \rho, \rho \rangle
$$
and
\beqa
\left(v'(x)-v'(x)^\ast\right) \dx \ceq \left( \begin{array}{cc} 0 & \pd_2 v_1(x) - \pd_1 v_2(x) \\ \pd_1 v_2(x) - \pd_2 v_1(x) & 0 \end{array} \right) {\dx_1 \choose \dx_2} \nonumber\\ \!\! & = & \!\! \left(\pd_1 v_2(x) - \pd_2 v_1(x)\right) {-\dx_2 \choose \dx_1}\nonumber\\ \!\! & = & \!\!  \left(\pd_1 v_2(x) - \pd_2 v_1(x)\right) |\dx|\, \rho^\bot .\label{skew}
\eeqa
The form (\ref{ee1}) of the Euler-Lagrange equation then follows on dividing (\ref{eeqn1}) by the expression
$1 - \alpha\, \langle v(x), \rho\rangle^2 + 2\alpha\, \langle v(x), \rho^\bot\rangle^2$
(which is strictly positive because $\alpha < 1$) and rearranging. \done

\bigskip\noindent
{\em Proof of Proposition \ref{p: rg}. }
Let us first state the Euler-Lagrange equation for the modified functional $x \mapsto \int_{t_0}^{t_1} \langle \dx(t), G(x(t))\, \dx(t) \rangle\, dt$; it is
\beq\label{modee}
0 = \ddot x + 2\alpha\, v\, \frac{d}{dt} \langle \dx,v \rangle + 2\alpha\, \langle \dx,v \rangle\, (v' - v'^\ast)\, \dx.
\eeq
(Here and in the following we suppress the argument $x$ of $v$ and $v'$, for compactness of notation.)
Forming the inner product with $\dx$ gives
\beqa\label{1stint}
0 \ceq \langle \dx, \ddot x \rangle + 2 \alpha\,  \langle \dx,v \rangle\, \frac{d}{dt} \langle \dx,v \rangle + 2\alpha\, \langle \dx,v \rangle\, \langle \dx,  (v' - v'^\ast)\, \dx \rangle\\ \ceq 
\frac{d}{dt} \left(\frac12\, |\dx|^2 + \alpha\, \langle \dx,v \rangle^2 \right) = \frac12\,\frac{d}{dt}\, \langle \dx(t), G(x(t))\, \dx(t) \rangle.\nonumber
\eeqa
Thus $\langle \dx(t), G(x(t))\, \dx(t) \rangle$ is constant as a function of $t$, or a ``first integral'', with the consequence that the Euler-Lagrange equation for the modified functional amounts to the same as the Euler-Lagrange equation for the original functional $x \mapsto L_G(x)$. Let us proceed with deriving the former equation.

With $\frac{d}{dt} \langle \dx,v \rangle = \langle \ddot x,v \rangle + \langle \dx,v' \dx \rangle$ and (\ref{skew}), equation (\ref{modee}) can be written as  
\beq\label{modee2}
0 = G \ddot x + 2\alpha \, \langle \dx,v' \dx \rangle\, v + 2 \alpha\,|\dx|\,  \langle \dx,v\rangle \left(\pd_1 v_2 - \pd_2 v_1\right) \rho^\bot,
\eeq
where again 
$$
G = I + 2 \alpha\, v \otimes v, \qquad \mbox{with inverse}\qquad G^{-1} = I - \frac{2\alpha}{1+ 2 \alpha} \, v \otimes v.
$$
Hence
$$
G^{-1} v = v/(1+2\alpha),\qquad G^{-1} \rho^\bot = \rho^\bot - \frac{2\alpha}{1+ 2 \alpha}\, \langle v,\rho^\bot \rangle\, v,
$$
so on writing $v = \langle v,\rho \rangle \rho + \langle v,\rho^\bot \rangle \rho^\bot$ and recalling the notation $\omega = \pd_1 v_2 - \pd_2 v_1$, we can state (\ref{modee2}) as a differential equation in explicit form,
\beqa
-\ddot x \ceq \frac{2\alpha|\dx|^2}{1+2\alpha}\, \bigg[\, \langle \rho,v' \rho\rangle \left(\langle v,\rho \rangle \rho + \langle v,\rho^\bot \rangle \rho^\bot \right) \nonumber\\ & & \qquad\ \,
+\, \langle v, \rho\rangle \, \omega \left(\rho^\bot (1 + 2\alpha) - 2\alpha\,  \langle v,\rho^\bot \rangle\, \{\langle v,\rho \rangle \rho + \langle v,\rho^\bot \rangle \rho^\bot \} \right) \bigg] 
\nonumber\\ \ceq
\frac{2\alpha|\dx|^2}{1+2\alpha}\, \bigg[\left( \langle \rho,v' \rho\rangle \langle v,\rho \rangle - 2\alpha\, \omega\, \langle v,\rho\rangle^2\, \langle v,\rho^\bot \rangle \right) \rho \nonumber\\ & & \qquad\ \,
+ \left(\langle \rho,v' \rho\rangle\, \langle v,\rho^\bot \rangle + \omega\, \langle v,\rho \rangle\ \{ 1+ 2\alpha\,\langle v,\rho\rangle^2\, \}\right) \rho^\bot \bigg]. \label{eexplct}
\eeqa
The proof of Proposition \ref{p: rg} is complete.
\done

\subsection*{A2. Approximations}

{\em Proof of Proposition \ref{p: picard}. }
Equation (\ref{eerg}) can be written as a first-order differential equation by means of the common recipe of enlarging the ``state space,'' from curves $x$ to pairs of curves $x, \dx$. The iteration (\ref{it1}), (\ref{it0}) then becomes the well-known Picard-Lindel\"of scheme, except that here we do not have an initial value problem for $x, \dx$; rather, the two endpoints of $x$ are fixed. There is only one obstacle for a straightforward application of the classical proof: one needs an a priori estimate for the distance of the iterates from the target, which has to remain bounded. Once this is achieved, it is a standard exercise to establish the boundedness and Lipschitz conditions necessary for an application of the Banach fixed point theorem. 

We leave that aside and concentrate on the a priori estimate. Let 
$$
J_{\alpha,n} = \int_{t_0}^{t_1} |\dx_{\alpha,n}(u)|^2\, du,
$$
and suppose initially that $\sup\nolimits_{\, \xi} \, ||v'(\xi)|| = M < \infty,$ the norm being declared as $||A|| = \sum_{j,k} |a_{j,k}|$ for matrices $A = (a_{j,k})$. Putting $\rho_0 = (\tau_1 -\tau_0)/\ell$ we have by (\ref{it1}) and (\ref{bndef})
\beq\label{it11}
\dx_{\alpha,n+1} = (\ell/T)\,\rho_0 + \alpha U_\alpha(x_{\alpha,n}, \dx_{\alpha,n})
\eeq
where
$$
U_\alpha(x, \dx)(t) =\int_{t_0}^t S_\alpha(x, \dx)(u)\, du - \frac{1}{T}\int_{t_0}^{t_1} \int_{t_0}^t S_\alpha(x, \dx)(u)\, du\, dt.
$$
From the straightforward bound
\beq\label{Sbound}
|S_\alpha(x,\dx)| \le 6\, |\dx|^2\, ||v'(x)||
\eeq
one readily gets the estimate 
\beq\label{Ubound}
||\,U_\alpha(x_{\alpha,n}, \dx_{\alpha,n})\,||_\infty \le 12 M J_{\alpha,n}.
\eeq
Thus by (\ref{it11})
\beqas
J_{\alpha,n+1} \ceq \int_{t_0}^{t_1} \bigg[(\ell/T)^2 + 2\alpha (\ell/T)\, \langle \rho_0,U_\alpha(x, \dx)(t)\rangle + \alpha^2\, |U_\alpha(x, \dx)(t)|^2 \bigg] \, dt\\ \cleq
\ell^2/T + 2\alpha \ell\, 12 M J_{\alpha,n} + T \left(\alpha12 M J_{\alpha,n}\right)^2\\ \ceq \ell^2/T\ \big[1 + \alpha 12 M J_{\alpha,n} T/\ell \big]^2,
\eeqas
or with $K_{\alpha,n+1} =J_{\alpha,n+1}\, T/\ell^2$ and $A_\alpha = \alpha 12 M \ell$,
$$
K_{\alpha,n+1}  \le \left(1 + \alpha 12 M K_{\alpha,n} (\ell^2/T)\, T/\ell \right)^2 = \left(1 + A_\alpha K_{\alpha,n} \right)^2.
$$
Since $J_{\alpha,0} = \int_{t_0}^{t_1} |\dot \tau(u) |^2\, du = \ell^2/T$, the starting value is $K_{\alpha,0} = 1$. 

We therefore have to study a ``sub-recursion'' of the form 
$$
x_{n+1} \le \varphi(x_n), \qquad \varphi(x) = (1+ax)^2, \qquad x_0 = 1,
$$
where $a$ is a positive constant. For $a< 1/4$ the function $\varphi$ has two fixed points, 
$$
z_\pm(a) = \left(1 - 2a \pm \sqrt{1-4a} \right)/(2a^2).
$$
As $a \downarrow 0$ the smaller fixed point remains bounded; in fact, $1 \le z_-(a) \le 4$ for all $a \in [0,1/4]$. We now proceed by induction.  We have $x_n \le z_-(a)$ for $n=0$, so suppose this holds for some $n\ge 0$. Then $\varphi(x_n) \le \varphi(z_-(a))$ by the monotonicity of $\varphi$, and hence $x_{n+1} \le \varphi(x_n) \le \varphi(z_-(a)) = z_-(a)$, as claimed.

The conclusion for our initial problem is that if we choose $\alpha^* < (48 M \ell)^{-1}$ then 
\beq\label{Jbound}
\sup\nolimits_{\, n \ge 0,\, \alpha \le \alpha^*}\, J_{\alpha,n} \le 4 \ell^2/T.
\eeq
As a consequence one has by (\ref{it11}) and (\ref{Ubound}) the uniform bound
\beq\label{dxbound}
||\dx_{\alpha,n+1}||_\infty \le \ell/T + \alpha\, 48 M \ell^2/T = \ell/T\left( 1 + \alpha\, 48 M \ell\right) \le 2 \ell/T;
\eeq
moreover, differentiating (\ref{it11}) and using (\ref{Sbound}) and (\ref{dxbound}) gives
$$
||\ddot x_{\alpha,n+1}||_\infty = \alpha\, ||S_\alpha(x_{\alpha,n},\dx_{\alpha,n})||_\infty \le \alpha\, 24 M \left(\ell/T\right)^2, 
$$
whence
\beq\label{dx2bound}
\int_{t_0}^{t_1}|\ddot x_{\alpha,n+1}(u)|\, du\le \alpha\, 24 M \ell^2/T \le \ell/(2T)
\eeq
(for every $n\ge 0$ and $\alpha \le \alpha^*$). The required a priori estimates are now obtained from (\ref{dx2bound}) by setting $y = \tau,\, x = x_{\alpha,n}$ in the following lemma, the easy proof of which is omitted.

\begin{lem} \label{itlem}
Suppose $x,\, y \in \X$ are such that $\int_{t_0}^{t_1}  |\ddot x(u) - \ddot y(u)|\, du \le B$. Then 
$$
||x-y||_\infty \le 2BT \quad \mbox{and} \quad ||\dx-\dot y||_\infty \le 2B.
$$
\end{lem}

Indeed, since $\ddot \tau = 0$ we may set $B=\alpha\, 24 M \ell^2/T$ and conclude that 
\beq\label{crucbnds}
||x_{\alpha,n}-\tau||_\infty \le \alpha\, 48 M \ell^2 \le \ell, \qquad ||\dx_{\alpha,n}-\dot \tau||_\infty \le \alpha\, 48 M \ell^2/T \le \ell/T
\eeq
for $n \ge 1,\, \alpha \le \alpha^*$. These estimates were derived under the assumption that $||v'(\xi)||$ is globally bounded. However, by (\ref{crucbnds}) and because $\alpha^*$ may be arbitrarily small, it suffices that $||v'(\xi)||$ is locally bounded in the vicinity of $\tau$, which it is. This concludes the crucial part of the proof. \done

\bigskip\noindent
{\em Proof of Proposition \ref{p: appsoln}. }
By definition we have $\ddot {\widehat x}_\alpha - \ddot x_{\alpha,1} = \alpha \left[S_0(\tau,\dot \tau) - S_\alpha(\tau,\dot \tau) \right]$, whence it readily follows that
$$
\int_{t_0}^{t_1}  |\ddot {\widehat x}_\alpha(u) - \ddot x_{\alpha,1}(u)|\, du = O(\alpha^2)
$$
(as $\alpha \to 0$). Therefore $||\widehat x_\alpha - x_{\alpha,1}||_\infty = O(\alpha^2)$, by Lemma \ref{itlem}, so applying (\ref{expconv}) with $n=1$ completes the proof of (\ref{gapp}). 

As for the second assertion, let us consider curves $y_\alpha \in \X$ with shape $\eta$ (i.e., of the form $y_\alpha = \tau + \alpha \eta + O(\alpha^2)$).
Let $\rho_\alpha \equiv \rho_{y_\alpha}  = \dot y_\alpha/ |\dot y_\alpha|$. Straightforward expansions give
\beqas
\rho_\alpha \ceq \rho_0 + \alpha (T/\ell) \langle \dot\eta, \rho_0^\bot \rangle \rho_0^\bot + O(\alpha^2),\\
\dot \rho_\alpha \ceq \alpha (T/\ell) \langle \ddot \eta, \rho_0^\bot \rangle \rho_0^\bot + O(\alpha^2).
\eeqas
The approximation 
$$
\rho_\alpha^\bot = \rho_0^\bot - \alpha (T/\ell) \langle \dot\eta, \rho_0^\bot \rangle  \rho_0 + O(\alpha^2)
$$
is readily verified on noting that $|\rho_0^\bot - \alpha (T/\ell) \langle \dot\eta, \rho_0^\bot \rangle  \rho_0|^2 = 1+ O(\alpha^2)$ and 
$$
\llangle \rho_0 + \alpha (T/\ell) \langle \dot\eta, \rho_0^\bot \rangle \rho_0^\bot,\ \rho_0^\bot - \alpha (T/\ell) \langle \dot\eta, \rho_0^\bot \rangle  \rho_0 \rrangle = O(\alpha^2).
$$
Thus for curves $y_\alpha$ with shape $\eta$ one has
\beq\label{rhoeta}
\langle \dot \rho_\alpha,  \rho_\alpha^\bot \rangle = \alpha (T/\ell)\, \langle \ddot \eta, \rho_0^\bot \rangle + O(\alpha^2).
\eeq
On the other hand, since $y_\alpha = \tau + O(\alpha)$ as $\alpha \to 0$, hence $|\dot y_\alpha| = \ell/T + O(\alpha)$, the right-hand side of (\ref{ee1}) evaluated at $y_\alpha$ behaves as
$$
-2\alpha\, (\ell/T)\ \big[ \langle v(\tau), \rho_0^\bot \rangle \langle \rho_0, v'(\tau) \rho_0 \rangle + \langle v(\tau), \rho_0 \rangle \, \omega(\tau) \big] + O(\alpha^2) = -2\alpha\, (\ell/T)\, \Fn_0(\tau) + O(\alpha^2).
$$
The comparison with (\ref{rhoeta}) shows that $y_\alpha$ satisfies the Euler-Lagrange equation (\ref{ee1}) up to terms of order $O(\alpha^2)$ if and only if 
$$
\langle \ddot \eta, \rho_0^\bot \rangle = -2 (\ell/T)^2\, \Fn_0(\tau).
$$
By (\ref{sgexpl}), this condition is equivalent to $\langle \ddot \eta, \rho_0^\bot \rangle = \langle \ddot \sigma, \rho_0^\bot \rangle$, and hence, by the boundary conditions, also equivalent to $\langle \eta, \rho_0^\bot \rangle = \langle \sigma, \rho_0^\bot \rangle$. 
\done

\subsection*{A3. Context representations}

Consider two functions $c$ and $\vartheta$ as introduced in Section \ref{exs}, along with the family of context curves $u \rightarrow C_\theta(u) = (u,c(u, \theta))$ parameterized by $\theta$. The tangent direction of such a curve at the point $C_\theta(u)$ is given by the unit vector
$$
v(C_\theta(u))  = \frac{(\,1, \pd_1 c(u,\theta)\,)}{\sqrt{1 + (\pd_1 c(u,\theta))^2}}\, .
$$
Since by assumption there exists for every $\xi=(\xi_1,\xi_2)$ in some planar region $\Xi$ a parameter $\theta = \vartheta(\xi_1,\xi_2)$ such that $c(\xi_1,\vartheta(\xi_1,\xi_2)) = \xi_2$, the points $C_\theta(u) = (u,c(u, \theta))$ fill the region $\Xi$ and we get a vector field $v$ on $\Xi$ by setting 
\beq \label{vfdef}
v(\xi) = \frac{\left(\ 1\, ,\, \pd_1 c(\xi_1,\vartheta(\xi_1,\xi_2))\ \right)}{\sqrt{1 + (\,\pd_1 c(\xi_1,\vartheta(\xi_1,\xi_2))\,)^2}}\, .
\eeq
Toward calculating the Jacobian of $v$ note first that because of $\xi_2 = c(\xi_1,\vartheta(\xi_1,\xi_2))$ we have 
$$
1 = \frac{\pd}{\pd \xi_2}\, \xi_2  =  \pd_2 c \cdot \pd_2 \vartheta, \qquad 0 = \frac{\pd}{\pd \xi_1}\, \xi_2  =  \pd_1 c + \pd_2 c \cdot \pd_1 \vartheta,
$$
hence 
$$ 
\pd_1 \vartheta = - \frac{\pd_1 c}{\pd_2 c}, \qquad \pd_2 \vartheta = \frac{1}{\pd_2 c}\, ,
$$ 
where here and below it is understood that $\pd_k \vartheta$ and $\pd_k c$ are evaluated at the arguments $\xi_1,\xi_2$ and $\xi_1,\vartheta(\xi_1,\xi_2)$, respectively. Similarly, 
$$
\frac{\pd}{\pd \xi_1}\, \pd_1 c = \pd_{11}^2 c + \pd_{12}^2 c\cdot \pd_1 \vartheta = \pd_{11}^2 c - \pd_{12}^2 c \cdot \frac{\pd_1 c}{\pd_2 c}, \qquad \frac{\pd}{\pd \xi_2}\, \pd_1 c = \pd_{12}^2 c\cdot \pd_2 \vartheta = \frac{\pd_{12}^2 c}{\pd_2 c}\, .
$$
With $\Fp = 1 + (\pd_1 c)^2$ the two components of $v$ can be written as $v_1 = \Fp^{-1/2},\, v_2= \Fp^{-1/2}\, \pd_1 c$, respectively. We have
\beqas
\pd_1 v_1 \!\!\! & = & \!\!\! \frac{\pd}{\pd \xi_1}\, \Fp^{-1/2} \, = \, -\Fp^{-3/2}\ \pd_1 c\, \left(\pd_{11}^2 c - \pd_{12}^2 c \cdot \frac{\pd_1 c}{\pd_2 c} \right),\\
\pd_2 v_1 \!\!\! & = & \!\!\! \frac{\pd}{\pd \xi_2}\, \Fp^{-1/2} \, = \, -\Fp^{-3/2}\ \pd_{12}^2 c \cdot \frac{\pd_1 c}{\pd_2 c}\, ,
\eeqas
furthermore
\beqas
\pd_1 v_2 \ceq \Fp^{-1/2} \left(\pd_{11}^2 c - \pd_{12}^2 c \cdot \frac{\pd_1 c}{\pd_2 c} \right) \, - \, \Fp^{-3/2}\, (\pd_1 c)^2 \left(\pd_{11}^2 c - \pd_{12}^2 c \cdot \frac{\pd_1 c}{\pd_2 c} \right)\\ 
\ceq \Fp^{-3/2}\left(\pd_{11}^2 c - \pd_{12}^2 c \cdot \frac{\pd_1 c}{\pd_2 c} \right),
\eeqas
and similarly
$$
\pd_2 v_2 \, = \, \Fp^{-3/2}\ \frac{\pd_{12}^2 c}{\pd_2 c}\, .
$$
The Jacobian of the vector field (\ref{vfdef}) thus is
\beq \label{jacob}
v'(\xi) \, = \, \Fp^{-3/2} \left( \begin{array}{cc} -\pd_1 c \left(\pd_{11}^2 c - \pd_{12}^2 c \cdot \frac{\pd_1 c}{\pd_2 c} \right) &  -\pd_{12}^2 c \cdot \frac{\pd_1 c}{\pd_2 c} \\ \pd_{11}^2 c - \pd_{12}^2 c \cdot \frac{\pd_1 c}{\pd_2 c}  & \frac{\pd_{12}^2 c}{\pd_2 c} \end{array} \right) , 
\eeq
and its rotation is 
\beq \label{rot}
\omega(\xi) = \Fp^{-3/2}\, \pd_{11}^2 c\, 
\eeq
(with the convention that all those partial derivatives are evaluated at $\xi_1,\vartheta(\xi_1,\xi_2)$).

Given a planar curve $x$ with associated 2-gon $\rho,\, \rho^\bot$, one readily derives explicit expressions for the crucial quantities $\Ft_\alpha(x),\, \Fn_\alpha(x)$ from (\ref{jacob}) and (\ref{rot}). For example, if as in Section \ref{exs} we take $x=\tau$ where $\tau(t) = (t,0),\, t \in [-\ell/2,\ell/2]$, the 2-gon is constant along $\tau$, with $\rho = (1,0),\, \rho^\bot=(0,1)$, and we get for $\alpha=0$
\beqas
\Ft_0(\tau) \ceq -\frac{\pd_1 c \left(\pd_{11}^2 c - \pd_{12}^2 c \cdot \frac{\pd_1 c}{\pd_2 c} \right)}{\left(1 + (\pd_1 c)^2\right)^2},\\
\Fn_0(\tau) \ceq \frac{\pd_{11}^2 c- \left(\pd_1 c\right)^2 \left(\pd_{11}^2 c - \pd_{12}^2 c \cdot \frac{\pd_1 c}{\pd_2 c} \right)}{\left(1 + (\pd_1 c)^2\right)^2}.
\eeqas
The latter expression gives (\ref{n0xpl}).

\subsection*{A4. Gaussian curvature}

{\em Derivation of the approximation (\ref{GK}) to the Gaussian curvature. } It is a consequence of Gauss's {\em theorema egregium} that the Gaussian curvature $K$ of the Riemannian geometry induced by the metric tensor $G$ can be expressed in terms of $G$ itself \cite{Lau65}. A formula convenient for our purpose is \cite[p.~114]{Str61}
\beq\label{GKexpl}
K = -\frac{1}{2 \sqrt{g}} \left\{ \pd_2 \left(\frac{\pd_2 E - \pd_1 F}{\sqrt{g}}\right) + \pd_1 \left(\frac{\pd_1 G - \pd_2 F}{\sqrt{g}} \right) \right\} - \frac{1}{4g^2} \left| \begin{array}{ccc} E & \pd_1 E & \pd_2 E \\ F & \pd_1 F & \pd_2 F \\ G & \pd_1 G & \pd_2 G  \end{array} \right|
\eeq
Here we have switched to the classical notation 
$$
E \equiv g_{11} = 1 + 2\alpha v_1^2, \quad F \equiv g_{12} = g_{21} = 2 \alpha v_1 v_2, \quad G \equiv g_{22} = 1 + 2\alpha v_2^2, \quad g \equiv g_{11} g_{22} - g_{12}^2,
$$
thereby accepting a change in the meaning of the symbol $G$, which will not cause confusion. Now $g = 1 + 2\alpha$ is constant (since $||v||=1)$. Therefore, $\sqrt{g}\, $ times the term in curly brackets on the right-hand side of (\ref{GKexpl}) is equal to
\beqas
\!\!\!\!\!\!\!\!\! && \pd_{22}^2 E + \pd_{11}^2 G - 2\pd_{12}^2 F = 2 \alpha \left\{ \pd_{22}^2 v_1^2 + \pd_{11}^2 v_2^2 - 2\pd_{12}^2 (v_1 v_2) \right\}\\ \ceq 4 \alpha \left\{ (\pd_2 v_1)^2 + v_1 \pd_{22}^2 v_1 + (\pd_1 v_2)^2 + v_2 \pd_{11}^2 v_2 - \pd_1 v_1 \pd_2 v_2 - v_1 \pd_{12}^2 v_2 - \pd_1 v_2 \pd_2 v_1 - v_2 \pd_{12}^2 v_1 \right\}\\ \ceq
4 \alpha \left\{ (\pd_1 v_2 - \pd_2 v_1)^2 + \pd_1 v_2 \pd_2 v_1 - \pd_1 v_1 \pd_2 v_2 + v_2\, \pd_1 (\pd_1 v_2 - \pd_2 v_1) - v_1\, \pd_2 (\pd_1 v_2 - \pd_2 v_1) \right\} \\ \ceq
4 \alpha \left\{ \omega^2 - v_1\, \pd_2 \omega + v_2\, \pd_1 \omega - (\pd_1 v_1 \pd_2 v_2 - \pd_1 v_2 \pd_2 v_1) \right\},
\eeqas
with $\omega$ the rotation of $v$. The last term in brackets, $\pd_1 v_1 \pd_2 v_2 - \pd_1 v_2 \pd_2 v_1$, equals the determinant of the Jacobian matrix $v'$. This determinant vanishes because $v'$ is rank-deficient due to the constraint $v_1^2+v_2^2=1$. (This can be seen also from (\ref{jacob}).) The determinant in the last term of (\ref{GKexpl}) is of the order $O(\alpha^2)$. Therefore, putting everything together one finds that 
\beqas
K \ceq - \frac{4\alpha}{2(1+2\alpha)} \left(\omega^2 - v_1\, \pd_2 \omega + v_2\, \pd_1 \omega\right) + O(\alpha^2) \\ \ceq
2\alpha \left(v_1\, \pd_2 \omega - v_2\, \pd_1 \omega - \omega^2 \right) + O(\alpha^2),
\eeqas
which is (\ref{GK}). \done

\begin{figure}[t]
\insfigc{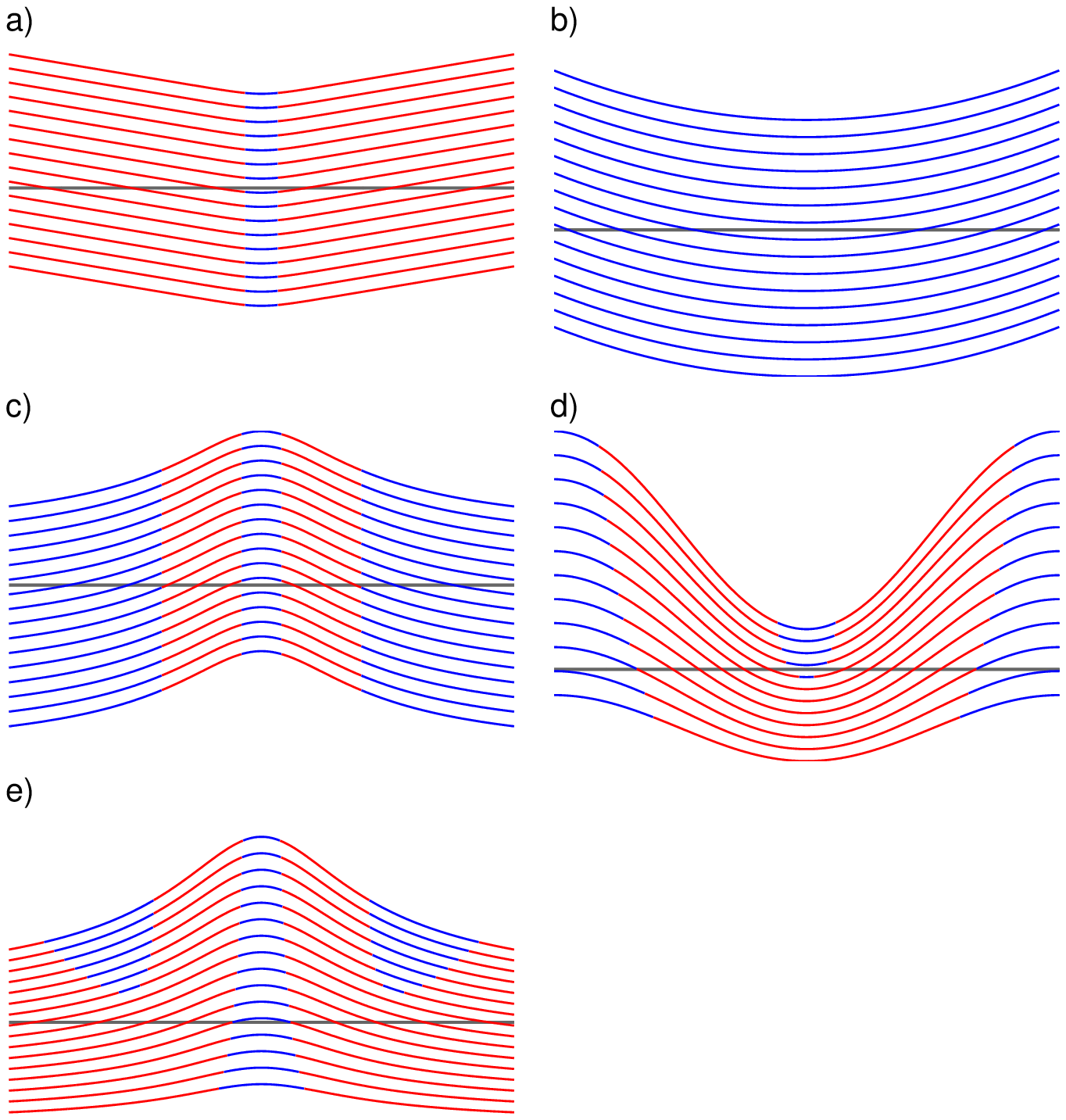}{0.875}

\fns
Figure 6:
Gaussian curvature of the Riemannian manifold associated with the respective context. Regions with (approximately) positive or negative curvature ($\FC > 0$ or $\FC <0$) are marked red and blue, respectively. Labels a) to e) refer to the same contexts as shown in Fig.~3. For details see Section \ref{Gaucurv}.\par
\end{figure}

%


Explicit expressions for the quantity $\FC = v_1\, \pd_2 \omega - v_2\, \pd_1 \omega - \omega^2$ can be derived from (\ref{rot}) for each of the three context types. In Fig.~6, parts of the respective context are color-marked depending on the sign of $\FC$. The vertical stripes in the panels a and c reflect the independence of $\FC$ on its second argument for contexts of type 1: $\FC = \FC(\xi_1)$ in this case. Such simplification does not occur with type 2 contexts (panels d, e). By contrast, $\FC = 0$ everywhere for contexts of type 3 (not shown). This could be verified using (\ref{rot}) and the particular form of the function $c$ in this case. It is more illuminating to recall that the context curves are (segments of) concentric circles, which by translation invariance can be assumed to be centered at the origin. The corresponding vector field then is $v(\xi) = (\xi_2,-\xi_1)/\sqrt{\xi_1^2 + \xi_2^2}$, from which $\FC=0$ easily follows.



\end{document}